%% file: Estimation_and_testing_for_multiple_regulation_of_multivariate_mixed_outcomes.tex
\documentclass[11pt]{article}
\usepackage{graphicx, amssymb, color, 
 latexsym,MnSymbol,soul, amsmath, mathrsfs}
\usepackage[a4paper, total={7in, 8in}]{geometry}
\input{GrandMacros}


\def\purple{\color{black}}

\definecolor{gray}{RGB}{150,  150,  150}
\definecolor{blue2}{RGB}{050,  050,  250}
\definecolor{cyan}{RGB}{024,  055,  155}
\definecolor{darkpink}{RGB}{255,  045,  180}
\definecolor{purple}{RGB}{255,  000,  250}
\def\red{\color{black}}
\def\blue{\color{black}}

\def\sm{^{(m)}}
\def\Vbb{\mathbb{ V}}
\def\yone{y^{(1)}}
\def\ym{y\sm}
\def\yM{y^{(M)}}
\def\bbetam{\bbeta\sm}
\def\hm{h\sm}
\def\epm{\epsilon\sm}
\def\Lscm{\Lsc\sm}
\def\betaz{\bbeta_0}

\def\smstar{^{*(m)}}
\def\sone{^{(1)}}
\def\stwo{^{(2)}}
\def\sM{^{(M)}}
\def\Rscr{\mathscr{R}}
\def\Pbb{\mathbb{ P}}
\def\summM{\sum_{m=1}^M}
\def\sumjp{\sum_{j=1}^p}
\def\dj{d_j}
\def\alphamj{\alpha_j\sm}
\def\betam{\beta\sm}
\def\betamj{\betam_j}
\def\alpham{\alpha\sm}
\def\Xbbtilde{\widetilde{\mathbb{X}}}
\def\ddLscm{\ddot{\Lsc}\sm}
\def\bLambdatildem{\bLambdatilde\sm}
\def\bvp{\boldsymbol{\varphi}}
\def\bvptilde{\widetilde{\bvp}}
\def\bSigma{\boldsymbol{\Sigma}}
\def\lamn{\lambda_n}
\def\gm{g^{(m)}}

\begin{document}
\title{Estimation and testing for multiple regulation of multivariate mixed outcomes}
\author{Denis Agniel, Katherine P. Liao, Tianxi Cai}


\maketitle

\begin{abstract}
Considerable interest has recently been focused on studying multiple phenotypes simultaneously in both epidemiological and genomic studies, either to capture the multidimensionality of complex disorders or to understand shared etiology of related disorders. We seek to identify {\em multiple regulators} or predictors that are associated with multiple outcomes when these outcomes may be measured on very different scales or composed of a mixture of continuous, binary, and not-fully-observed elements. We first propose an estimation technique to put all effects on similar scales, and we induce sparsity on the estimated effects. We provide standard asymptotic results for this estimator and show that resampling can be used to quantify uncertainty in finite samples. We finally provide a multiple testing procedure which can be geared specifically to the types of multiple regulators of interest, and we establish that, under standard regularity conditions, the familywise error rate will approach 0 as sample size diverges.
Simulation results indicate that our approach can improve over unregularized methods both in reducing bias in estimation and improving power for testing.
\end{abstract}


\section{Introduction}

Considerable recent interest has been focused on studying multiple phenotypes simultaneously in both
epidemiological and genomic studies. There are several reasons for such
studies to be important. First, a complex disorder is usually associated with multiple correlated phenotypes.
Hence, even when the focus of the study is on a single disease, multiple phenotypes might be needed to fully capture the complexity
and multidimensionality of the disorder. Second, multiple related disorders might share the same
etiology and a joint assessment will enable researchers to identify factors associated with risk of multiple diseases.
As an example, recent studies have identified common genes associated with a higher risk of what were previously considered distinct autoimmune diseases \cite{zhernakova2009detecting}.
Similar shared genetic bases have also been suggested for various types of cancers and related psychiatric disorders \cite{solovieff2013pleiotropy}.
Identification of predictors of multiple outcomes, also commonly known as multiple {\em traits} in the genetics literature, can improve understanding of disease etiology, genetic regulatory pathways, and treatment. Further complicating matters, the outcome measures may be {\em diverse}: they may be binary (e.g., presence of disease), continuous (disease activity score), ordinal (severity of disease), not completely observable (perhaps due to a limit of quantification), or any combination thereof.

To address these questions statistically,  we seek to assess the association between a vector of predictors $\bx=(x_1, ..., x_p)\trans$ and a vector of outcomes $\by=(y^{(1)}, ..., y^{(M)})\trans$ by estimating and testing all relevant effects. For each predictor $x_j$ we desire an estimation and testing procedure that will identify its associated subset of $\by$. In particular, researchers often want to identify predictors that are important for multiple or all outcomes. We will call $x_j$ a ``multiple regulator" if it is associated with multiple outcomes, a terminology which we adapt from \cite{peng2010regularized}. An example of what we call multiple regulation is known as pleiotropy in the genetics literature.
Our goal of identifying multiple regulation is not to be confused with identifying predictors that are associated with any outcomes. Association with any outcomes is an active area of research, with two examples being global association tests and group-sparse regularization. Global tests provide a test for the relationship between $x_j$ and the entire set $\by$ \cite{jiang1995multiple, he2013general}
 and have been shown in some situations to have higher power than marginal tests to detect associations when $x_j$ relates to multiple outcomes
. Group-sparse methods, largely based on the group lasso \cite{yuan2006model}, use model selection to identify predictors that are relevant for any outcome \cite{turlach2005simultaneous}. These methods, while powerful and useful, do not address the question of \emph{which} outcomes are relevant for each predictor and in general are unsuited for diverse outcomes that may contain censoring.

Here, we are particularly interested in identifying predictors that are relevant for multiple outcomes and inferring which subset of $\by$ each of the $x_j$'s are associated with. There is a paucity of literature that addresses these specific questions. Under linear regression models, the remMap procedure \cite{peng2010regularized} addresses such a question via variable selection by jointly penalizing both the $L_1$ and $L_2$ group norms of a squared loss. Under generalized linear models, one could potentially modify the hierarchical lasso \cite{zhou2010group} procedure, originally proposed to handle grouped predictors with a single outcome, to address the multiple regulator problem. 
 When making joint inference on a diverse set of outcomes, it is also desirable to put all effects on similar scales. A simple example of this idea can be found in \cite{schifano2013genome}, where linear regression models were considered for multiple continuous outcomes and each outcome was scaled by its standard deviation. 
However, none of these methods is applicable to settings where $\by$ consists of a diverse set of outcomes whose scales may not be easily comparable to each other, 
{\red especially when $\by$ may contain censored time-to-event variables.} 
To accommodate modeling of multiple outcomes of different scales and/or type,  we propose in this paper the use of semiparametric transformation models 
which give all effects of $\bx$ on $\by$ a similar interpretation. A liability thresholding version of such models can naturally model binary or ordinal outcomes.

Regardless of estimation technique, a multiple testing procedure is required to control error rates when identifying multiple regulation, which operates on the (potentially large) set of hypotheses $\{H_j\sm: x_j$ unassociated with $y\sm\}_{j = 1, ..., p; m = 1, ..., M}$. Neither \cite{peng2010regularized} nor \cite{zhou2010group} tackles this issue. In general, multiple testing based on regularized estimation is challenging for two reasons. First, while many of the regularization procedures such as \cite{zhou2010group} established asymptotic {\em oracle properties} for their estimators --- non-informative predictors can be detected with no uncertainty and their detection induces no additional variation in the estimation of the informative predictors  \cite{fan2001variable,zou2006adaptive} --- in finite samples those properties may be far from holding. Consequently, basing testing procedures on such asymptotic results may lead to inflated type I error in finite samples. Second, the
estimators and hence their corresponding test statistics could be highly correlated from the regression fitting. Standard methods for controlling the familywise error rate (FWER), like the Bonferroni procedure, tend to be conservative in the presence of correlation, and they ignore the dependence structure in the data.

\begin{figure}
\begin{center}
\begin{tabular}{c}
\includegraphics[width = \textwidth, keepaspectratio]{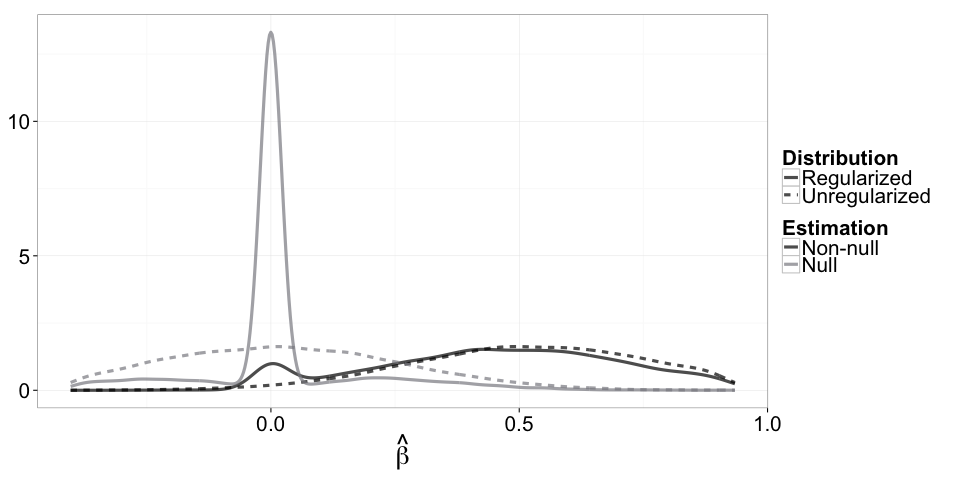}
\end{tabular}
\end{center}\vspace{0.1in}
\caption{Sampling distributions of null and non-null effects, with and without regularization. Tails of the distributions are truncated for ease of presentation.}\label{sampdist}
\end{figure}

We propose a two-stage technique to both estimate the effects of $\bx$ on $\by$ and identify multiple regulation while controlling error rates. In the first stage, we posit models to put all effects on the same scale, and we use regularization to induce sparsity in the estimated effects. To do this, we generalize the adaptive hierarchical lasso of \cite{zhou2010group} to handle the case of semiparametric models. In the second stage, we
employ a stepdown procedure analogous to \cite{romano2005exact} to identify multiple regulation while controlling error rates. Our two-stage method, entitled Sparse Multiple Regulation Testing (SMRT), is powerful for several reasons. First, our modeling strategy allows us to do estimation and make inference on outcomes that may be measured on completely different scales. Next, regularization enables us to more efficiently estimate both the null and non-null effects. The null effects are estimated as 0 with probability tending to 1 and the non-null effects are estimated with lower variability compared to unregularized estimators. Furthermore, the distributions of the estimates of null effects and the distributions of the estimates of non-null effects are distinctly separated through regularization, giving us more power to detect the non-null effects (see figure \ref{sampdist} for an illustration from our simulations). Finally, our testing procedure can be specifically geared to detect associations with multiple outcomes.

However, it is generally challenging to perform testing based on regularized estimators since their distributions in finite samples cannot be approximated well by asymptotic results. We lay out
permutation- and resampling-based procedures to better approximate the finite-sample distributions of the proposed test statistics and the regression parameter estimators.
This enables us to properly control error rates for both hypothesis testing and interval estimation. {\purple Thus, in addition to providing the estimator $\bbetahat$ based on joint regularization, the main contributions of this paper include providing resampling procedures to make joint inference about $\bbetahat$ and deriving the SMRT testing procedure to identify the subset of outcomes associated with each of the predictors. Our proposed estimation and testing procedures can account for the joint effects of the predictors and the correlation among both the predictors and the outcomes.}

The rest of the paper is organized as follows. In section 2, we give an overview of SMRT. In section 3, we discuss details regarding our sparse estimator, including its asymptotic properties and quantifying its variability. In section 4, we discuss issues related to testing, including the asymptotic guarantee of familywise error control and practical approaches to finite-sample error control. In section 5, we apply our method to a genetic study of autoantibodies with the goal of identifying multiple regulators of autoimmunity. Simulation results which validate our method are provided in section 6. And finally, in section 7, we discuss implications and further directions. 

\section{Overview of SMRT}\label{testing}
Suppose the data for analysis consists of  $n$ independent and identically distributed random vectors $\Vbb = \left\lbrace \bV_i = (\by_i\trans, \bx_i\trans)\trans\right\rbrace_{ i = 1, ..., n}$ where $\by_i = (\yone_i, ..., \yM_i)\trans$ are the $M$ outcomes and $\bx_i = (x_{i1}, ..., x_{ip})\trans$ are the $p$ predictors for the $i$th subject. 
We first propose a unified modeling strategy for diverse $\by$ by assuming that
\begin{align}
P(\ym \le y \mid \bx) = \gm\{\bx\trans\bbeta_0\sm + h\sm(y)\} ,
	\qquad m = 1, ..., M, \label{model}
\end{align}

\noindent where $\bbeta_0\sm$ represents the unknown effect of $\bx$ on $y\sm$, $\hm(\cdot)$ is an unspecified smooth, increasing function, and the link function, $\gm$, is given although the correlation structure of $\by$ is left unspecified.
For ease of presentation, we assume that $\by$ is fully observed although the proposed method can easily accommodate censored outcomes. When $\ym$ is continuous, (\ref{model}) is equivalent to
\begin{align}
h\sm(\ym) = - \bx\trans\bbeta_0\sm + \epm, \quad \mbox{with } P(\epm \le z \mid \bx) = P(\epm \le z) = \gm(z) . \label{model2}
\end{align}

\noindent Generalized linear models for a binary or ordinal outcome can be written in the form of (\ref{model}) and (\ref{model2}) by viewing the observed outcome as a thresholded version of a latent continuous outcome and $h\sm$ as only defined at the threshold values, as previously suggested in the literature \cite[e.g.]{thomas1998mixed}.
Choice of $\gm$ determines the type of model being fit. For example, $\gm(x) = e^x/(1+e^x)$ corresponds to a proportional odds model for continuous $\ym$
and a logistic regression model if $\ym$ is binary. Models (\ref{model}) and (\ref{model2}) have also been previously used to analyze censored survival outcomes \cite{cai2000semiparametric,zeng2007maximum}.
The virtue of this approach is that the scale of the $\bbeta\sm$ will be comparable across $m = 1, ..., M$ {\purple when the same or comparable $\gm(x)$ are used} whether $y\sm$ is continuous, discrete, or not fully observed because each marginal model has a similar form. {\blue For example if $\gm(x) = e^x/(1+e^x)$, then each $\beta_j\sm$ has the interpretation of a log odds ratio regardless of whether $y\sm$ is continuous, binary, ordinal, or censored.}

To estimate $\bbeta_0\sm$, one may employ the non-parametric maximum likelihood estimator (NPMLE)  under model (\ref{model})  \cite{zeng2007maximum,
murphy2000profile} based on data observed on the $m$th outcome, $\Vbb\sm = \{(\ym_i, \bx_i\trans)\trans\}_{i = 1, ...,n}$. Let $\Lscm(\bbeta\sm)$ denote the resulting profile log-likelihood (PLL) function corresponding to the NPMLE.
It has been shown that under mild smoothness conditions, the profile likelihood can be treated as a regular likelihood, and the maximum PLL estimator $\bbetatilde\sm = \argmax{\bbeta\sm}{\Lscm(\bbeta\sm)}$ is regular and semiparametric efficient \cite{murphy2000profile}. However, when $p$ is not too small and  $\bbeta_0 = (\bbeta_0^{(1)\transpose}, ..., \bbeta_0^{(M)\transpose})\trans$ might be sparse, an improved estimator may be obtained by imposing regularization on the PLL. To do this, we simultaneously consider all $M$ outcomes and obtain a sparse $\bbetahat = (\bbetahat^{(1)\transpose}, ..., \bbetahat^{(M)\transpose})\trans$ as the minimizer of the penalized sum of negative PLLs
\begin{align}
	-\summM \Lscm(\bbetam) + p_{\lambda, \bw}(\bbeta) \label{garrote}
\end{align}
with penalty function
$p_{\lambda, \bw}(\bbeta) = \sumjp \dj + \lambda \summM\sumjp w_j\sm\left|\alphamj\right|, \text{ with } \betamj = d_j \alphamj, \text{ subject to }d_j \ge 0.$
The penalty function $p_{\lambda, \bw}(\cdot)$ was previously proposed in \cite{zhou2010group} for generalized linear models with grouped predictor variables. The tuning parameter $\lambda$ controls the amount of regularization and weight $w_j\sm = |\betatilde_j\sm|\inv$ is 
chosen to ensure oracle properties of $\bbetahat$. Summing over the PLLs in (\ref{garrote}) essentially imposes a working independence assumption across the outcomes \cite{liang1986longitudinal}. 
Imposing the joint penalty $p_{\lambda, \bw}(\bbeta)$ incorporates the potential for joint sparsity across all outcomes for some $x_j$'s.
Setting $d_j = 0$ declares $x_j$ to be non-informative for all outcomes or equivalently $\bbeta_{0j} = (\beta_{0j}^{(1)}, ..., \beta_{0j}^{(M)})\trans=0$; while setting
$\alphamj = 0$ suggests that $\beta_{0j}\sm = 0$.
We will show that $\bbetahat$ possesses a  {\em sparsistency} property, i.e., $P(\betahat_j\sm = 0 |  \beta_{0j}\sm = 0) \rightarrow 1$. This ensures desirable asymptotic properties for our testing procedures. We give further details regarding $\bbetahat$ and its asymptotic properties in section \ref{sec-estimation}. We now turn to the topic of testing.

\subsection{Testing a single predictor $x_j$}\label{single_test}
In order to make inference on a single predictor, SMRT employs a stepdown procedure for $x_j$ considering the $M$ hypotheses $\Hsc_j = \{H_j\sm : \beta_{0j}\sm = 0\}_{m = 1, ..., M}$ with alternative hypotheses denoted
$\{\Hbar_j\sm : \beta_{0j}\sm \neq 0\}_{m = 1, ..., M}$.
To test $H_j\sm$, we consider the statistic $t_j\sm = \nhalf\left|\betahat_j\sm\right|/\sigmatilde_j\sm$ and its reference distribution $\Tsc_j\sm = \{t_j^{*_b(m)}\}_{b = 1, ..., B}$
which approximates the distribution of $t_j\sm \mid H_j\sm$ and can be obtained by, for example, resampling or permutation (see section \ref{refdist}). We scale $\betahat_j\sm$ by $\sigmatilde_j\sm$, which is an estimated standard error of $\nhalf(\betatilde_j\sm - \beta_{0j}\sm)$, since under $H_j\sm$, $\sigmahat_j\sm= \widehat{\var}\{\nhalf(\betahat_j\sm-\beta_{0j}\sm)\}^{1/2} \to 0$ and the null distribution of $\nhalf\betahat_j\sm/\sigmahat_j\sm$ is difficult to approximate.

To test $\Hsc_j$ simultaneously, we order the test statistics  $\bt_j = (t_j\sone, ..., t_j\sM)\trans$ from largest to smallest, $
t_j^{(r_1)} \geq t_j^{(r_2)} \geq ... \geq t_j^{(r_M)},
$
and identify their corresponding hypotheses $H_j^{(r_1)}, ..., H_j^{(r_M)}$. 
Define for every ${\Omega} \subset \{1, ..., M\}$ the sup-statistic over $\Omega$ and its corresponding reference distribution: $s_j^{\Omega} = \max_{m \in {\Omega}} t_j\sm$ and $\Ssc_j^{\Omega} = \{\max_{m \in {\Omega}} t_j^{*_b(m)}\}_{b = 1, .., B}$. Furthermore, denote the $\psi$th quantile of $\Ssc_j^{\Omega}$ by $c_j^{\Omega}(\psi)$, which approximates the $\psi$th quantile of $s_j^{\Omega}$ under the null that $\{\beta_{j}\sm = 0: m \in \Omega\}$.
We identify the subset of hypotheses to reject, denoted by $\Rscr_j$, as follows. 
\begin{enumerate}
\item[1)] Let ${\Omega}_1 = \{1, ..., M\}$. If $s_j^{\Omega_1} \leq c_j^{{\Omega}_1}(\psi)$, accept all hypotheses and stop. Otherwise, let $\Rscr_j = \{r_1\}$ and continue. ...\\
\item[l)] Let $\Omega_l = \Omega_1 \setminus \Rscr_j$. If $s_j^{\Omega_l} \le c_j^{\Omega_l}(\psi)$, accept all hypotheses in $\{H_j^{\sm}\}_{m \in \Omega_l}$ and stop. Otherwise, let $\Rscr_j = \Rscr_j \cup \{r_l\}$  and continue. ...\\
\item[M)] Let $\Omega_M = \{r_M\}$. If $s_j^{\Omega_M} \leq c_j^{r_M}(\psi)$, accept $H_j^{(r_M)}$. Otherwise, let $\Rscr_j = \Rscr_j \cup \{r_M\}$.
\end{enumerate}

The stepdown procedure for the simultaneous testing of $\Hsc_j$ then rejects all hypotheses in $\{H_j^{(m)}\}_{m \in \Rscr_j}$ and concludes that $x_j$ is associated with $\{y\sm\}_{m \in \Rscr_j}$. If the reference distribution and $\psi$ are chosen such that the probability of making a type I error at each step is at most $\alpha$: 
\begin{align}
P\left(\left.s_j^{\Omega_k} > c_j^{\Omega_k}(\psi) \right| \bigcap_{m \in \Omega_k}H_j^{\sm}\right) \leq \alpha,\label{tI}
\end{align}
for any $k$, then the FWER of the stepdown procedure -- that is, the probability of making at least one false rejection over the set $\Hsc_j$ -- is maintained at $\alpha$. We discuss in detail issues relating to the choice of reference distribution and $\psi$ in section \ref{testing_prop}. We also describe how, regardless of the choices of reference distribution and $\psi$, the FWER is asymptotically 0 because $\bbetahat$ is sparsistent.

\subsection{Multiple regulation testing}
Now suppose scientific interest lies only with a predictor if it regulates at least $k$ outcomes. That is, we only care to conclude that $x_j$ is associated with $\{y\sm\}_{m \in \Rsc_j}$, for some $\Rscr_j\subset \{1, ..., M\}$ if the number of rejections (i.e., the cardinality of $\Rscr_j$) is at least $k$. Then we can modify the testing procedure in the previous section to increase power to detect $k$-multiple regulators (kMRs) at the expense of being able to detect if $x_j$ appears to be associated with fewer than $k$ outcomes. The testing procedure proceeds by essentially skipping the first $k-1$ steps in the previous section and only rejecting the first $k-1$ hypotheses if any other hypotheses are rejected. Thus, we will either reject 0 hypotheses or $k$ or more hypotheses. Throughout, when we refer to SMRT, we mean the combination of our sparse estimation technique and our multiple regulation testing procedure for a given $k$, with $k = 1$ corresponding to the application of the test in the previous section. 

We identify the subset of hypotheses to reject, denoted by $\Rscr_j$, as follows: 
1) let $\Omega_1 = \{r_{k} , ..., r_M\}$. If $s_j^{\Omega_1} \le c_j^{\Omega_1}(\psi)$, accept all hypotheses and stop. Otherwise, let $\Rscr_j = \{r_1, ..., r_k\}$  and continue; 
2) let $\Omega_2 = \{1, ..., M\} \setminus \Rscr_j$. If $s_j^{\Omega_2} \le c_j^{\Omega_2}(\psi)$, accept all hypotheses in $\{H_j^{\sm}\}_{m \in \Omega_2}$ and stop. Otherwise, let $\Rscr_j = \Rscr_j \cup \{r_{k+1}\}$  and continue.
Steps 3 through $M-k+1$ proceed as in the previous section.

{\red As discussed in section \ref{testing_prop}, the stepdown test with $k > 1$  also has asymptotic FWER of 0. In addition to requiring \eqref{tI}, which we will call controlling the common type I error, we also require the control of a second type of error: {\em incorrectly} rejecting one of $\{H_j\sm\}_{m = r_1, ..., r_{k-1}}$ based on {\em correctly} rejecting $H_j^{(r_k)}$ in step one. We will call this a type I error by implication.
Since the distribution of null effects gets shrunk dramatically toward 0 (see figure \ref{sampdist}), it is unlikely for this type of error to occur in practice because it requires a test statistic corresponding to a null hypothesis to be larger than a test statistic from a rejected alternative hypothesis.}
{\blue We leave discussion of controlling the FWER for all predictors to appendix \ref{allpredix}. The extension of the testing procedure for a single predictor is straightforward. } 

\section{Inference about $\bbetahat$}\label{sec-estimation}
We next detail the construction of $\bbetahat$ as well as the asymptotic distribution for the zero and non-zero components, which is crucial for the validity of our estimator, confidence intervals, and proposed testing procedures. 
Estimation proceeds by minimizing \eqref{garrote}. Now, since the profile log-likelihoods $\{\Lscm\}_{ m = 1, ..., M}$ are non-linear functions without closed form in most cases, direct maximization of (\ref{garrote}) may be numerically challenging, especially when $p$ is not small. To reduce the computational complexity and enable the use of widely available software, we propose to take a quadratic expansion of $\Lscm(\bbeta\sm)$ in (\ref{garrote}) similar to \cite{zhang2007adaptive} and \cite{ wang2007unified}. Specifically, we instead minimize
\begin{align}
\| \bYtilde - \Xbbtilde \bbeta\|_2^2 + p_{\lambda, \bw}(\bbeta),\label{penlik1}
\end{align}
where $\bItilde\sm = -\ddLscm(\bbetatilde\sm)$, $\ddLscm(\bb) = \partial^2 \Lscm(\bb)/\partial \bb\partial\bb\trans$,
 $\Xbbtilde =  \text{diag}(\bLambdatilde\sone, ..., \bLambdatilde\sM)$,
 $\bYtilde= \Xbbtilde\bbetatilde$
and $\bLambdatildem$ is a symmetric half matrix of $\bItilde\sm$ such that $\bItilde\sm = \bLambdatildem\bLambdatildem$. 
Computational simplifications and a full algorithm for fitting are discussed in appendix \ref{algorithm}.

\subsection{Asymptotic Theory}
In this section, we present the properties of our proposed estimator $\bbetahat$. It has the property of {\em sparsistency} in that it asymptotically sets truly null effects to exactly 0. Specifically, define $\Asc$ and $\Asc^c$ as indexing the non-zero and zero components of $\betaz$, respectively, where $\bbeta_{\Asc}$  denotes the subvector of $\bbeta$ corresponding to $\Asc$. Then a {\em sparsistent} estimator $\bbetahat$ is one that satisfies $P(\bbetahat_{\Asc^c} = 0) \rightarrow 1$  as $n \rightarrow \infty$. Furthermore, our estimates of non-null effects are asymptotically normal and possess the {\em oracle property}, in that they are as efficient in the limit as if we knew which effects were truly null \emph{a priori}. Let $\bI_{\Asc,\Bsc}$ denotes the submatrix of $\bI$ corresponding to rows in $\Asc$ and columns in $\Bsc$.

In appendix \ref{asymp}, we show that for PLLs $\{\Lscm(\bbetam)\}_{m = 1, ..., M}$ that satisfy certain regularity conditions (listed in appendix \ref{appint}), if $n\inv\sqrt{\lambda} = o_p(n^{-1/2})$, then there exists a root-$n$ consistent local maximizer $\bbetahat$ such that $P(\bbetahat_{ {\Asc^c}} = 0) \rightarrow 1$ and
$\nhalf(\bbetahat_{\Asc} - \bbeta_{0\Asc}) \to N(0,\bI_{\Asc,\Asc}^{-1}\bSigma_{\Asc,\Asc}\bI_{\Asc,\Asc}^{-1})$ in distribution,
where $\bSigma_{\Asc,\Asc} = \cov(\bvp_{i\Asc}(\bbeta_0))$, $\bvp_{i\Asc}(\bbeta_\Asc)$ denotes the contribution of the $i$th subject to the profile score function for $\bbeta_{\Asc}$, $\bI = \diag\{\bI\sone, ..., \bI\sm\}$, and $\bI\sm$ is the limiting information matrix. 
This result, parallel to that given in  \cite{zhou2010group}, offers the promise of identifying null effects with probability approaching 1, while efficiently estimating non-null effects. From a testing perspective, it ensures that the type I error of SMRT for any $k$ decreases to 0 as $n \rightarrow \infty$.

\subsection{Estimating the variability in $\bbetahat$}\label{resamp}
The asymptotic results on $\bbetahat$ suggest that we are as efficient in the limit as if we knew which parameters were truly 0 from the outset. However, in finite samples the added variability due to estimating ${\Asc^c}$ may not be negligible, and hence relying on the asymptotic result will underestimate the variability in $\bbetahat$. To better approximate the finite-sample  distribution, we propose a perturbation resampling procedure to estimate the distribution of  $\nhalf(\bbetahat - \bbeta_0)$. This procedure, by accounting for the variability in estimating ${\Asc^c}$, provides a more precise estimate of the variability in $\bbetahat$ and maintains the correlation structure in $\bbetahat$.

We generate a resampled counterpart of $\bbetahat$, denoted by $\bbetahat^*$, in two steps. We first generate $\bbetatilde^*$, a resampled version of $\bbetatilde$, by either perturbing the profile likelihood or directly perturbing the influence function corresponding to $\bbetatilde$. {\blue In essence, each perturbation is achieved by multiplying $G_i$ to the likelihood contribution from the $i$th subject,  where the positive perturbation variables $\{G_i\}$ are generated independently with mean 1 and variance 1. Then we minimize our objective function (\ref{penlik1}) using $\bbetatilde^*$ in place of $\bbetatilde$, yielding resampled estimates $\bbetahat^*$.} 
Similar resampling procedures have been proposed for making inference with a wide range of standard objective functions without regularization \cite[e.g]{tian2007model,uno2007evaluating}
and recently extended to accommodate $L_1$-type regularized estimators \cite{minnier2011perturbation}. Here, we propose such a resampling procedure to both account for the potential correlation among the outcomes and better approximate the finite-sample behavior of hierarchically regularized estimators. 

In appendix \ref{app_ptb}, we {\red detail the perturbation procedure and establish its asymptotic properties, which are parallel to those for $\bbetahat$. A key feature of the resampled $\bbetahat^*$ is that } $\nhalf\left(\bbetahat^*_\Asc - \bbetahat_\Asc\right) \mid \Vbb$ has the same limiting distribution as $\nhalf(\bbetahat_\Asc - \bbeta_{0\Asc})$. Thus, to approximate the distribution of $\bbetahat$ for a given dataset, we may generate a large number of $\bbetahat^*$s, $\left\{\bbetahat^{*_b}\right\}_{b=1,...,B}$ for some suitably large $B$. To construct a confidence interval (CI) for a specific $\betamj$, one may estimate the standard error of $\betahat_j^{(m)}$ as $\sigmahat_j^{(m)}$ the empirical standard error of its perturbed realizations, $\left\lbrace\betahat_{j}^{*_b(m)}\right\rbrace_{b=1,...,B}$. A $100(1-\alpha)\%$ level confidence interval can then be constructed based on the normal confidence interval $\betahat_j^{(m)} \pm \Zsc_{1-\alpha/2} \sigmahat_j^{(m)}$ or alternatively the lower and upper $\alpha/2$ percentiles of $\left\lbrace\betahat_{j}^{*_b(m)}\right\rbrace_{b=1,...,B}$. 

\subsection{Tuning}
SMRT involves a large number of minimizations and tuning parameter selections.
It is thus not feasible to select $\lambda$ using time-consuming methods such as cross-validation. We propose a modified BIC criteria: $\lambda = \argmin{\lambda} \left(\| \bYtilde - \Xbbtilde \bbeta_\lambda\|_2^2 + \min\{n^{0.1}, \log n\}\text{df}_\lambda/n\right),$ where $\bbeta_\lambda$ is the minimizer of (\ref{penlik1}) corresponding to $\lambda$ and df$_\lambda$ is the number of non-zero entries in $\bbeta_\lambda$. In small and moderate sample sizes, $n^{0.1}$ is much smaller than $\log n$ and is used here. However, when $n$ becomes large $\log n$ may be preferred. {\blue  \cite{wang2007unified} showed that this BIC criteria (with either $\log n$ or $n^{0.1}$) satisfies the rate requirements} {\purple for a standard adaptive LASSO type penalty. Similar arguments can be used to justify the rate for the adaptive hierarchical LASSO type penalty used in (\ref{penlik1}).}

\section{Testing}\label{testing_prop}
In this section, we show that the FWER of our testing procedure is asymptotically 0 because of the sparsistency of $\bbetahat$. We also discuss in more detail the choice of reference distribution.

\subsection{Properties of SMRT}\label{sec-smrt-fwer}
One of the main results of this paper is that, given a suitably estimated $\bbetahat$, the FWER of our stepdown procedure approaches 0 as $n \rightarrow \infty$ for any $k$ regardless of the reference distribution or what quantile $\psi$ we use to determine the cutoff for rejection. Specifically, we show in appendix \ref{step} that
if $\bbetahat$ is sparsistent, then for every $j$ and $\Omega$, $
P\left(\left.s_j^{\Omega} > c_j^{\Omega}(\psi) \right| \bigcap_{m \in \Omega}H_j^{\sm} \right) \rightarrow 0$
as $n \rightarrow \infty$, and SMRT has an asymptotic FWER of 0, for any reference distribution, $k$, and $\psi$.
The result follows from showing that common type I errors and type I errors by implication both occur with probability tending to 0. With regard to common type I errors, under a given null $H_j\sm$, the test statistic $t_j\sm$ is estimated at exactly 0 with probability tending to 1 and, under the composite null $\bigcap_{m \in \Omega} H_j\sm$, $s_j^{\Omega}$ tends to 0 as well. Thus, we cannot reject $\bigcap_{m \in \Omega} H_j\sm$, regardless of the value of $c_j^{\Omega}(\psi)$, and therefore common type I errors will occur with probability approaching 0 as $n \rightarrow \infty$. The other potential source of type I error occurs for $k > 1$ when incorrectly rejecting $H_j^{(r_{k'})}$ based on correctly rejecting $H_j^{(r_{k})}$, $k' < k$. However, this sort of type I error will only occur if the test statistic for a null hypothesis {\red (which is tending to 0)} is larger in magnitude than the test statistic for an alternative hypothesis{\red, which is of course impossible asymptotically. }

While the foregoing result shows that the asymptotic behavior of SMRT is ensured by the sparsistency of $\bbetahat$, of course in finite samples choice of reference distribution and $\psi$ is paramount in maintaining the desired error rate. Maintaining the FWER at approximately $\alpha$ can be ensured by choosing the reference distribution and $\psi$ such that the probability of making a type I error at each step of the testing procedure is maintained at approximately $\alpha$.

\subsection{Choosing a reference distribution}\label{refdist}
\begin{figure}
\begin{center}
\begin{tabular}{c}
\includegraphics[width = \textwidth, keepaspectratio]{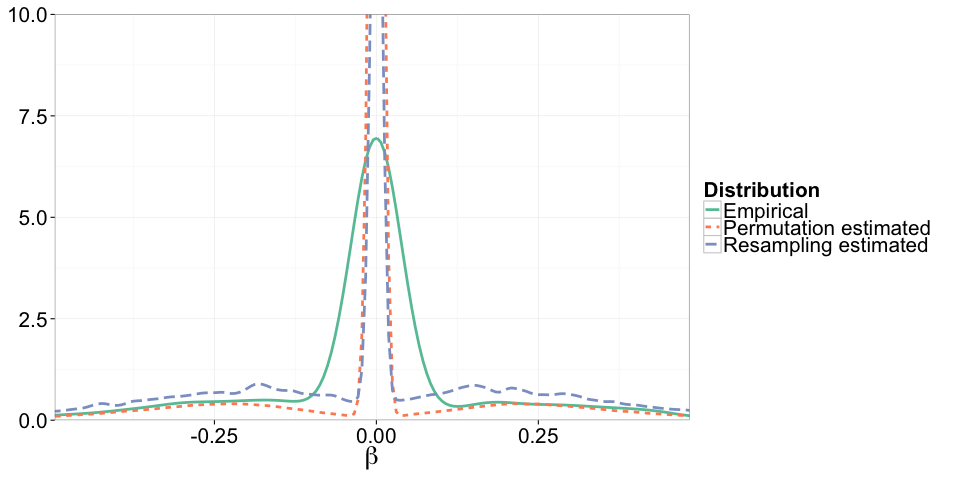}
\end{tabular}
\caption{Simulation-based empirical and estimated distribution of null effects. Empirical null distribution of $\betahat_j\sm$ (labelled "Empirical") agrees closely in the tails with the permutation-based estimate (labelled "Permutation estimated"), while the resampling-based estimate (labelled "Resampling estimated") overestimates the density in the tails.}\label{ptbpmt}
\end{center}
\end{figure}

As discussed in the previous section, any reference distribution $\Tsc_j\sm = \{t_j^{*_b(m)}\}_{b = 1, ..., B}$ will provide asymptotic control of the FWER by virtue of sparsistency of the estimator $\bbetahat$. {\red We explore resampling- and permutation-based reference distributions. The resampling-based reference distribution is based on $t_j^{*_b(m)} = \nhalf\left|\betahat_j^{*_b(m)} - \betahat_j\sm\right|/\sigmatilde_j\sm$. Simulation results suggest that, although resampling provides good approximation to the finite-sample distribution of $\bbetahat_{\Asc}$, it tends to over-estimate the variability of $\bbetahat_{\Asc^c}$ (see figure \ref{ptbpmt}). As an alternative, we consider a permutation-based reference distribution with $t_j^{*_b(m)}$ based on an estimate of $\bbeta_0$ from a dataset where $y\sm$ is permuted. See appendix \ref{app_ref} for further details about the procedure and section \ref{sec-sim} for simulation results.
Numerical results suggest that,the permutation-based reference distribution does a better job of approximating the finite-sample null distribution of $t_j\sm$, as shown in figure \ref{ptbpmt}.}

\subsection{Choosing $\psi$}
To control the FWER at $\alpha$-level, it seems reasonable to choose $\psi = 1 - \alpha$. This ensures that, for a suitable reference distribution and $n$ large enough, $P\left(\left.s_j^{\Omega} > c_j^{\Omega}(\psi) \right| \bigcap_{m \in \Omega} H_j^{\sm}\right) \lessapprox \alpha$, for any $\Omega$, which, along with negligible type I error by implication, will give approximate FWER control. 
In light of the fact that all type I errors are tending to probability 0, one could obtain improved power by choosing $\psi < 1- \alpha$, while maintaining the level at $\alpha$. This is particularly important if using the resampling-based reference distribution. One could use another layer of permutation or resampling to estimate the smallest $\psi$ that would still maintain the level $\alpha$. However, that requires computing a large number of permutations/resamples for each of the $B$ members of the reference distribution, which becomes prohibitively computationally demanding quickly. Computing a suitable $\psi < 1 - \alpha$ is a topic of future research.

\section{Genetic study to identify shared autoimmune risk loci}\label{sec-example}
We apply SMRT to a study of shared autoimmunity with the goal of identifying genetic markers associated with 4 autoantibodies: anti-nuclear antibodies (ANA),  anti-cyclic citrullinated protein (CCP) antibodies, anti-transglutaminase (TTG) antibodies, and anti-thyroid peroxidase antibodies (TPO). These 4 autoantibodies are respectively markers for 4 autoimmune diseases (ADs): systemic lupus erythematosus (SLE), rheumatoid arthritis (RA), celiac disease, and autoimmune thyroid disease. The genetic markers consists of 67 single-nucleotide polymorphisms (SNPs) previously published as potential risk markers for these four ADs.  Discovering which SNPs regulate multiple ADs can aid in understanding potential shared pathways or etiology of these diseases \cite{zhernakova2009detecting}. While it is rare for an individual to have multiple ADs, 
multiple autoantibodies can be present in individuals predisposed to having the multiple ADs even in the absence of the disease phenotypes. Here we consider the autoantibodies markers for subjects at higher risk for the ADs.

The study cohort includes 1265 individuals of European ancestry with RA identified through electronic medical records at Partners Healthcare \cite{liao2010electronic}.
 Due to a limit of quantification, the antibody measurements are highly unreliable when the values are either very low or very high. A convenient approach to incorporating such limitations is by assuming a marginal proportional odds model and truncating the observations at the limit of quantification. Hence $\beta_{0j}\sm$ still has the interpretation of being a log odds ratio (OR).

\begin{figure}
\begin{tabular}{cc}
(a) & (b)\\\includegraphics[height = \textheight, width = 0.5\textwidth, keepaspectratio]{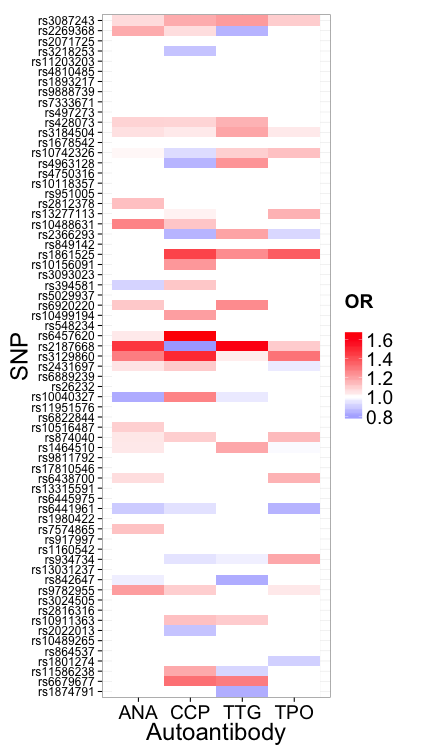} &\includegraphics[height = \textheight, width = 0.5\textwidth, keepaspectratio]{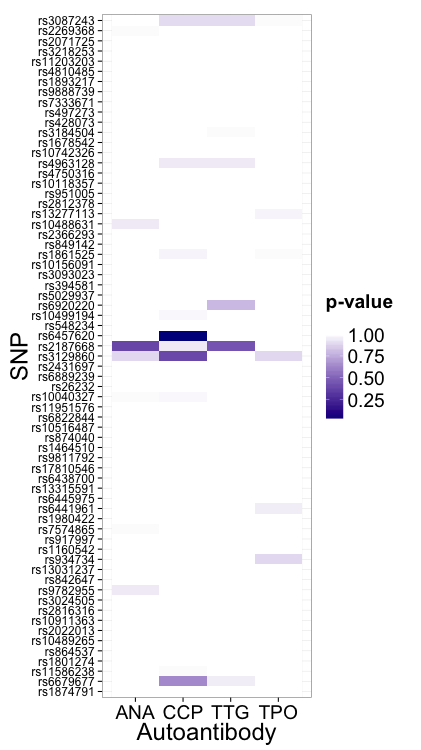}
\end{tabular}
\caption{Results for autoantibody data. SNPs are listed on the $y$-axis, and autoantibodies are listed on the $x$-axis. (a) Sparse effect estimates. Darker colors indicate larger magnitudes, and white indicates no estimated association. (b) Adjusted $p$-values. Darker color indicates smaller $p$-value and more evidence against the null hypothesis of no association.}\label{aab_plot}
\end{figure}

Results for the autoantibody data are summarized in figure \ref{aab_plot}. Figure \ref{aab_plot} (a) shows results for the sparse estimation step. In the figure, SNPs are denoted along the $y$-axis, and outcomes are denoted along the $x$-axis. Color of the tile indicates the OR estimate, with darker colors indicating stronger association.
In order to measure the strength of association with respect to the FWER, we provide adjusted $p$-values as the smallest $\alpha$ such that that test would reject while controlling the FWER for the SNP at $\alpha$. Figure \ref{aab_plot} (b) shows this $p$-value for each test.

Due to the large number of hypotheses, we do not have sufficient power to detect multiple regulation while simultaneously controlling the FWER across all SNPs. Taking a less conservative view, if we control the FWER at the SNP level, five SNPs show some evidence of multiple regulation at $\alpha = 0.1$. The two strongest associations were with rs2187668 and rs3129860. Having previously shown associations to SLE \cite{taylor2011risk} and celiac \cite{van2007genome}, rs2187668 was estimated to be related to the autoantibodies for those diseases at OR = 1.45 ($p$-value 0.005) for ANA and OR = 1.62 ($p$-value 0.005) for TTG, as well as to CCP (OR = 0.78, $p$-value 0.05). This SNP is in the MHC region, which is known to affect immune function. Similarly, rs3129860, also in the MHC region, which had previously shown an association to SLE \cite{taylor2011risk}, here demonstrated an association to ANA (OR = 1.28, $p$-value 0.05) , CCP (OR = 1.50, $p$-value 0.003), and TPO (OR = 1.30, $p$-value 0.05). 
\section{Simulation results}\label{sec-sim}
We ran simulations to assess the performance of our point and interval estimation procedures as well as SMRT. We loosely based our simulations on the autoantibody dataset, allowing the relationship between $\bx$ and $\by$ to be specified by a proportional odds model. We considered sample sizes of 150, 250, and 500 and ran 1000 simulations for each sample size. For each simulation, 1000 resampled $\bbetahat^*$s were generated.

We set the number of predictors of interest $p$ to be 30 and the number of outcomes $M$ to be 4. Covariates $\bx$ took values in $\{0, 1, 2\}$ with probability $\{p^2, 2p(1-p), (1-p)^2\}$ where $p = 0.15$. Outcomes $\by$ were generated according to the marginal proportional odds model, conditional on $\bx$. We allowed correlation in $\by$, which was accomplished by first generating correlated normal random variables $\bz_i \sim N_4(\bzero, \Sigma)$ where $\Sigma = 0.85I + 0.15\bone\bone\trans$ is exchangeable. Then let $\bu_i = \Phi(\bz_i)$ for Gaussian distribution function  $\Phi(\cdot)$, and finally $\by_i = \exp(\bx_i\bbeta_0 + \bepsilon_i)$ where $\bepsilon_i = \log(\frac{u_i}{1-u_i}) \sim $ logistic. For computational simplicity, we discretized $\by$ into ten levels {\blue of roughly equal sizes according to deciles. The only change when discretizing is to the number of locations at which $h\sm$ is estimated. In practice, this is not an issue (note that we did not discretize in the data analysis), but for the purposes of simulation it was a moderate speed-up with little information loss.}

The relationship between $\bx$ and $\by$ is defined by
\begin{align*}
	(\bbeta_0\sone, ...,\bbeta_0\sM)_{30 \times 4}
	 = \left(\begin{matrix}
	\bone_{20} & {\bf\frac{1}{2}}_{16} & \bone_{12} & {\bf\frac{1}{2}}_8\\
	\bzero_{10} & \bzero_{14} & \bzero_{18} & \bzero_{22}
	\end{matrix}\right)_{30 \times 4}.
\end{align*}
where $\bone_k$ is a $k\times 1$ vector of ones, $\bzero_k = 0\times \bone_k$ and ${\bf\frac{1}{2}}_{k} = \frac{1}{2}\times\bone_k$. This configuration indicates that there are eight predictors related to all four outcomes, four related to just the first three outcomes, four related to just the first two outcomes, and four related to just the first outcome. The remaining ten predictors are null, unrelated to any outcome. We also see that associations to outcomes $y^{(2)}$ and $y^{(4)}$ are weak, so we would expect there to be less power to detect those effects.

\subsection{Estimation}
\begin{figure}
\begin{center}
\begin{tabular}{c}
\includegraphics[width = \textwidth, keepaspectratio]{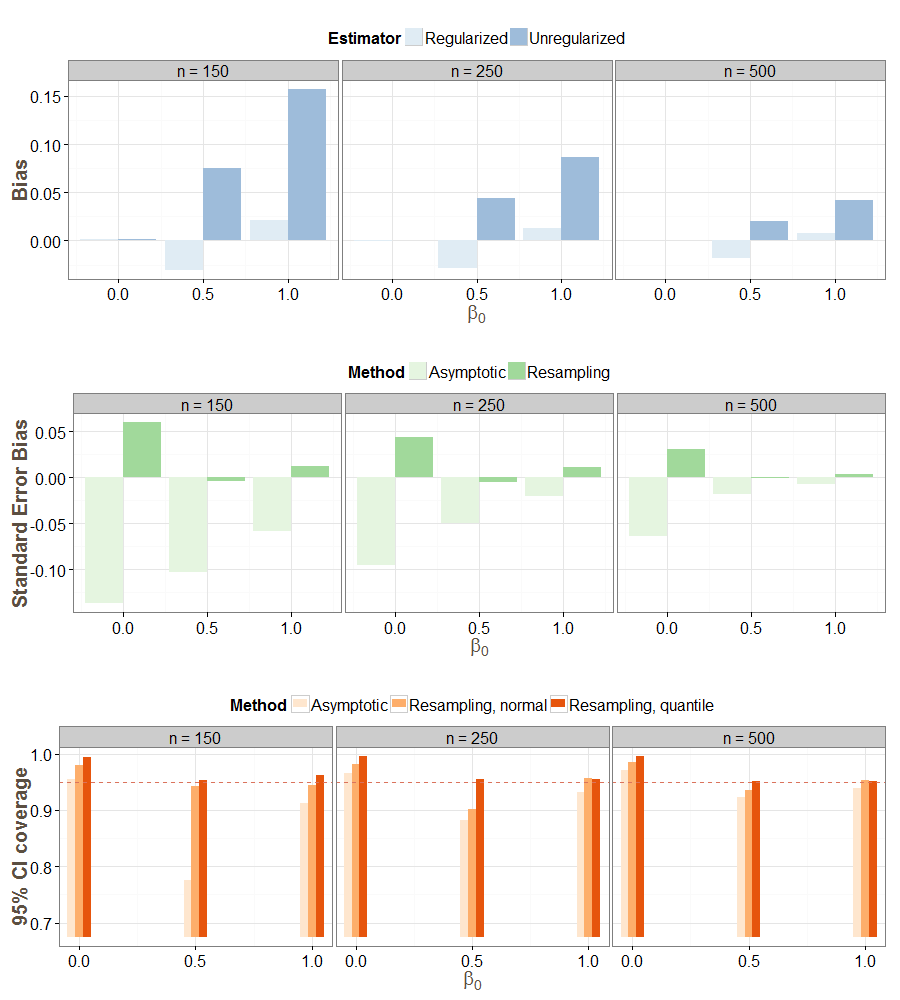}
\end{tabular}
\end{center}
\caption{Average performance of point estimates, standard errors, and confidence intervals across 1000 simulations at sample sizes of $n = 150, 250, 500$. All quantities are aggregated over $\beta_{0j}\sm$ and plotted against $\beta_{0j}\sm$. Top panel: Average estimated bias in regularized $\betahat_j\sm$ and unregularized $\betatilde_j\sm$. Middle panel: Average estimated bias of estimates of $\sigma\sm_j$, the standard error of $\betahat_j\sm$, comparing resampling-based estimates to asymptotic estimates. Bottom panel: 95\% CI coverage comparing asymptotic, resampling-based normal, and resampling-based quantile CIs.}\label{sims_plot}
\end{figure}

We first demonstrate that our point and interval estimation procedures perform well in finite samples. Figure \ref{sims_plot} (top panel) shows the average bias in $\bbetahat$ and $\bbetatilde$ across simulations, plotted according to true effect size $\bbeta_0$ and sample size. The regularized $\bbetahat$ exhibits much smaller bias than the unregularized $\bbetatilde$ for all sample sizes and effect sizes. Particularly at smaller sample sizes, regularization substantially reduces the bias in the estimator.

In figure \ref{sims_plot} (middle panel), we plot the average bias in SE estimates obtained based on our proposed resampling procedures as well as those based on the asymptotic variance. Both the asymptotic SE estimate and the resampling-based one $\sigmahat_j\sm$ overestimate the variability in $\betahat_j\sm$ when $\beta_{0j}\sm = 0$, but $\sigmahat_j\sm$ more closely approximates $\sigma_j\sm$. When $\beta_{0j}\sm  \neq 0$, the asymptotic SE tends to underestimate the true variability, while $\sigmahat_j\sm$ approximates it well.

We examine CI coverage in the bottom panel of figure \ref{sims_plot} and see that underestimating the SEs leads to poor 95\% CI coverage levels for the normal-based CI methods, based on $\sigmatilde_j\sm$ and $\sigmahat_j\sm$. Resampling-based quantile 95\% CIs have good coverage for all values of $\beta_{0j}\sm$ and all sample sizes. The coverage levels of asymptotic-based CIs are as low as 78\% for non-zero effects and remain lower than the nominal level even when $n = 500$. Hence in practice, we recommend the quantile-based CIs.

\subsection{Testing}
In the following sections, we examine the performance of SMRT. We first characterize the performance of our procedure with $k = 1$ when testing is performed for each predictor individually 
considering both the resampling-based  and the permutation-based reference distribution. Then we consider testing with $k > 1$ and controlling error rates for all predictors. 
\subsubsection{Resampling-based reference distribution}
We briefly demonstrate the gains in power possible by using the resampling-based reference distribution. For ease of presentation, we demonstrate the performance of the testing procedure for the marginal test of $H_j\sm$ with and without regularization. Results for the full stepdown procedure are similar.
\begin{figure}
\begin{center}
\begin{tabular}{c}
\includegraphics[width = 0.75\textwidth, keepaspectratio]{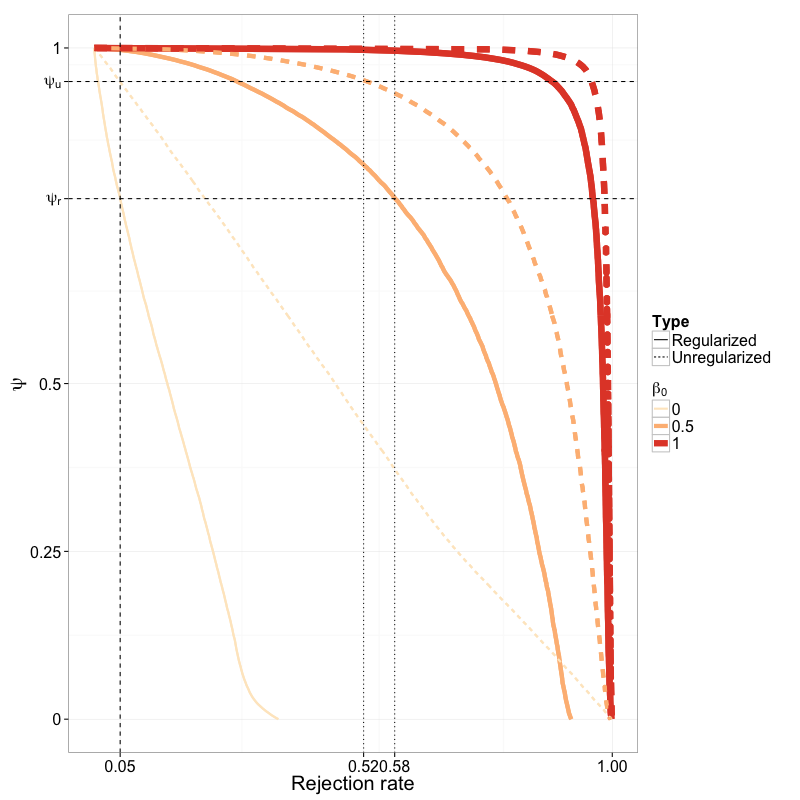}
\end{tabular}
\end{center}
\caption{Threshold $\psi$ ($y$-axis) plotted against its associated empirical rejection rate ($x$-axis) for the marginal test of $H_j\sm$ across 1000 simulations, with color denoting the magnitude of $\beta_{j0}\sm$ and linetype indicating whether regularization was used. The value on the $y$-axis $\psi_r$ represents the threshold at which the empirical type I error was controlled for the regularized test, and $\psi_u$ represents the threshold at which the empirical type I error was controlled for the unregularized test. Results for effects of the same magnitude are averaged for ease of presentation. }\label{ptb_rej_plot}
\end{figure}
Figure \ref{ptb_rej_plot} demonstrates the power gain possible when using the regularized estimator with the resampling-based reference distribution.  The plot shows the threshold necessary to obtain a given rejection rate. The ideal threshold maintains the rejection rate for null effects $(\beta_{0j}\sm = 0$) at a given level, say $\alpha = 0.05$, indicated by the vertical dashed line. That threshold that maintains the type I error for the regularized estimator, indicated by $\psi_r$ in the plot, is much lower than the threshold for the unregularized estimator, indicated by $\psi_u = 1 - \alpha = 0.95$ in the plot. Furthermore, the power to detect weak effects ($\beta_{0j}\sm = 0.5$) using the regularized estimator at $\psi_r$ is {\red 58\% compared to 52\% using the unregularized estimator at $\psi_u$}, while the power to detect strong effects are similar. Thus, if one could select $\psi_r$ adaptively, it appears that large power gains could be observed by using regularization. Due to its computational burden, however, we did not pursue this method further in our simulations.

\subsubsection{Permutation-based reference distribution}
We pursue a more rigorous study of SMRT using the permutation-based reference distribution {\red mentioned in section \ref{refdist} and described in detail in appendix \ref{app_ref}}. To demonstrate the role of regularization in improving testing, we compare SMRT to an identical testing procedure based on the unregularized $\bbetatilde$, named MRT. We use the permutation-based reference distribution for both SMRT and MRT and take $\psi = 1-\alpha$. To demonstrate the advantages of the stepdown method, we compare to a single-step procedure, denoted as Sup, where we reject all $H_j\sm$ for which $t_j\sm > c_j^{\Omega_1}(\psi)$ where $\Omega_1 = \{1, ..., M\}$. Finally, we compare to the Bonferroni adjustment. 

When controlling the FWER at $\alpha = 0.05$ for each $x_j$ using the basic test, SMRT and MRT performed similarly in controlling FWER. The average empirical FWER was .046, .052, and .055 at $n = 150, 250, 500$ respectively for SMRT. The corresponding average FWER for MRT was 0.042, 0.049, 0.054, at those respective sample sizes. The more conservative Sup test had average FWERs of .041, .044, and .043, respectively, and the even more conservative Bonferroni .028, .026, .021. 

\begin{figure}
\begin{tabular}{cc}
&\includegraphics[width = 0.33\textwidth, keepaspectratio]{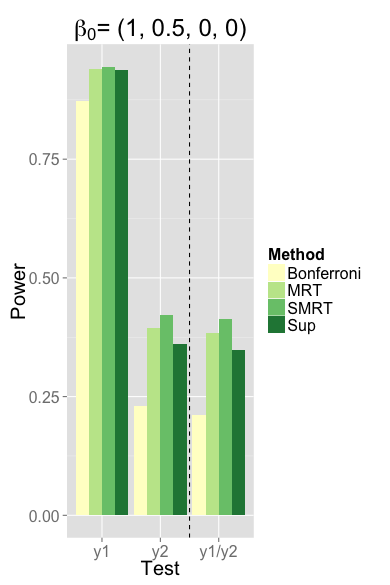}
\includegraphics[width = 0.67\textwidth, keepaspectratio]{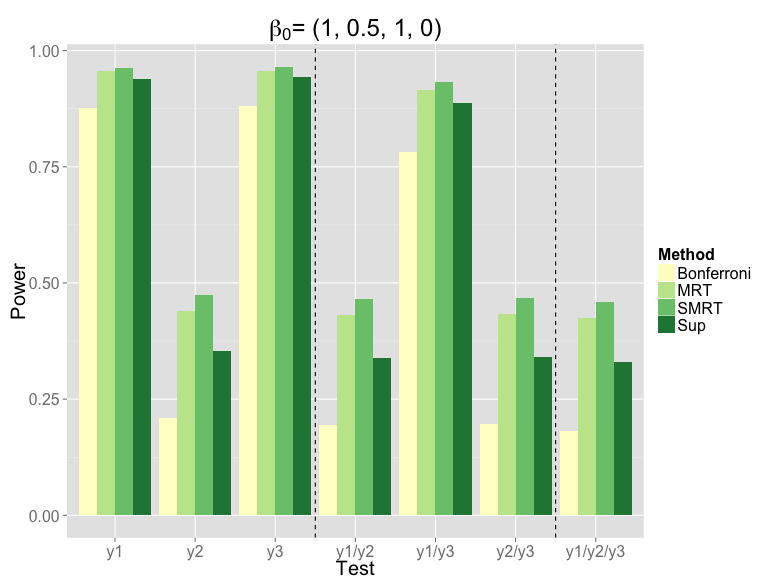}\\
&\includegraphics[width = \textwidth, keepaspectratio]{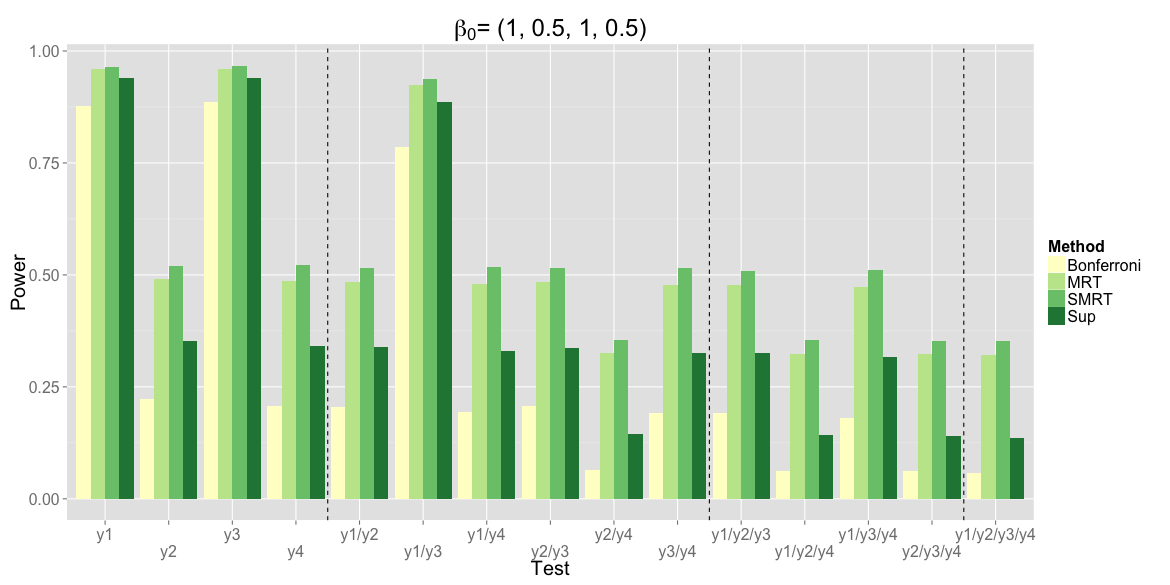}
\end{tabular}

\caption{Power to detect non-null effects across 1000 simulations at sample size $n = 250$ and level $\alpha = 0.05$. Each plot indicates how many outcomes the predictors tested are associated with. For example, the top left plot corresponds to predictors with strong association to $y\sone$, $\beta_{0j}\sone = 1$ and weak association to $y\stwo$, $\beta_{0j}\stwo = 0.5$. Tests are listed on the $x$-axis. Power is indicated on the $y$-axis. Power estimates are aggregated over all estimates that share the same effect sizes. To take a couple of examples, the bar corresponding to "$y1$" in the figure corresponds to power to reject $H_j\sone$, and the bar corresponding to "$y1/y2/y3$" in the figure corresponds to power to reject each of $H_j\sone, H_j\stwo, H_j^{(3)}$ simultaneously.}\label{rej_plot}
\end{figure}

In terms of power, SMRT dominates all other test procedures. Figure \ref{rej_plot} depicts the power to detect non-null effects at $n = 250$ (other sample sizes show similar relative performances, with SMRT performing relatively better as sample size decreases). Possible rejections are listed across the bottom, and results are arranged according to how many outcomes the predictor is actually associated with. The figure shows that SMRT is uniformly more powerful than MRT, Bonferroni and Sup, with the differences becoming more apparent in identifying multiple regulation.

Results for controlling the FWER by applying SMRT to all predictors and all hypotheses were qualitatively similar. All methods maintained the nominal level of the test, and SMRT obtained higher power than MRT, Sup, and Bonferroni at all sample sizes.
When we apply SMRT to each $x_j$ with $k = 2$, the average FWER for SMRT decreases to .020, .027, and .033 at $n = 150, 250, 500$. Taking $k = 3$ or $k=4$ sees a further reduction to FWERs. 

\subsection{Superiority of joint analysis over marginal models}
{\blue In this section, we demonstrate the advantages of performing a joint analysis for the detection of multiple regulation. We compare our estimator $\bbetahat$ to the estimator obtained by fitting each marginal model individually using $L_1$ penalization, which we will denote $\bbeta^\dag$.} 
 {\purple The joint analysis improves our ability to detect multiple regulation, with the improvement over $\bbeta^\dag$ increasing with the number of outcomes a predictor is associated with. 
{\red For example, when $n=500$, for the eight predictors associated with all four outcomes, the power to detect association with {\red $y^{(4)}$ and $y^{(2)}$ increased from 57\% to 67\% and from 60\% to 66\%, respectively, when using SMRT based on the joint penalty as opposed to marginal models, with a negligible increase for $y^{(3)}$ and $y^{(1)}$}. For predictors of three outcomes, SMRT based on $\bbetahat$ increased the power for association with $y^{(2)}$ to 61\% from 57\% for $\bbeta^{\dag}$. For predictors of two outcomes, $\bbetahat$ had 58\% power in detecting effects associated with $y\stwo$ compared to 54\% for $\bbeta^\dag$.}
Furthermore, $\bbetahat$ is much better at eliminating {\blue completely non-informative} predictors by estimating all of their effects at exactly 0. Using joint estimation, $\bbetahat$ eliminates null predictors completely 52\% of the time, while the rate is only 23\% using marginal models, when $n = 500$. {\blue The relative performance patterns} are similar for $n = 150$ and 250. }

\section{Discussion}\label{discussion}
We have proposed a framework for testing and estimation across a diverse set of outcomes, with the explicit goal of identifying predictors for multiple outcomes. This framework allows the combination of information across continuous, semi-continuous, and discrete outcomes while maintaining control of the FWER. We have extended existing sparse regression methods for identifying multiple regulation to the complex scenario when the components of $\by$ may be on very different scales or not completely observed. We have proven the asymptotic properties of this estimator and shown that one can use resampling to estimate its variability. We have, finally, provided a testing framework for identifying multiple regulation and demonstrated that the properties of the estimator ensure that the testing procedure has asymptotic FWER of 0.

While we rely on the sparsistency properties of our estimator, other penalty functions could potentially accomplish similar results to the hierarchical penalty we proposed. As long as sparsistency holds and a suitable finite-sample reference distribution can be obtained, e.g. through permutation or resampling, other penalty functions could be worth exploring.
For simplicity, we used a working independence assumption to combine the profile log-likelihoods of multiple outcomes. But when the outcomes are not independent, incorporating information about the covariance in $\by$ can improve efficiency \cite{liang1986longitudinal}. A further advantage to using the quadratic approximation to $\Lsc\sm$ in (\ref{penlik1}) instead of the profile log-likelihood itself (besides computational tractability) is that we can incorporate covariance information about $\by$ through the initial estimate $\bbetatilde$. If the (unpenalized) initial estimate $\bbetatilde$ is estimated in a way that gains efficiency by taking correlation in $\by$ into account, then that increase in efficiency will be propagated into our estimation of $\bbetahat$.

{\blue Due to the fine-grained nature of multiple regulation analysis and the complexity of dealing with diverse $\by$, SMRT may not be preferred in genome-wide or other very high-dimensional data where discovery is of primary importance. Rather than using SMRT to discover novel markers, we suggest using it to validate known markers. Global tests for all outcomes \cite{jiang1995multiple, he2013general} provide better power to discover unknown risk markers. Theoretically, the convergence of our estimators cannot be guaranteed jointly unless the number of predictors $p$ and outcomes $M$ is finite. Thus, we  require $M$ not to be too large compared to the sample size. Practically speaking, the computational complexity of the estimation procedure grows with $M$. A brief simulation yielded average run times of 0.4,1.7, 4.3, 11.0, 21.2, 44.9, and 343.8 seconds for $M = 4, 8, 12, 16, 20, 25,$ and 50, respectively, at $n = 500$, with results being quite similar with $n = 150$.}

Finally, we have focused on FWER as the error rate of primary interest throughout this paper, but it could be of interest in some testing situations to employ less restrictive error control, especially when the number of tests grows large and signals are weak. We could easily extend SMRT with $k=1$ to include more generalized error rates, such as $k$-FWER or the false discovery proportion, as in \cite{romano2010balanced}, and testing when $k > 1$ could be adapted in that direction as well. 

\appendix
\section{Appendix}\label{appint}
For the following proofs, we put mild restrictions on the model \eqref{model}
 for each outcome $y\sm$, as described in section 3 of \cite{murphy2000profile}. We reproduce the restrictions here for completeness. Since the requirements hold for each outcome, we drop the superscripts $\sm$ for the moment. Let $\log l(\bbeta, h)(x)$ be the full log-likelihood. We require that there exists a $h_t(\bbeta, h)$ such that $\ell(t, \bbeta, h)(x) = \log l(t, h_t(\bbeta, h))(x)$ is twice continuously differentiable for all $x$ with first and second derivatives denoted $\dot{\ell}(t, \bbeta, h)(x)$ and $\ddot{\ell}(t, \bbeta, h)(x)$. Further, $h_{\bbeta}(\bbeta, h) = h, \text{for every } (\bbeta, h).$ And $\dot{\ell}(\bbeta_0, \bbeta_0, h_0)$ must be the efficient score function. For every fixed $\bbeta$, let $\hhat_{\bbeta}$ be the NPMLE for $h$. Then, for any $\bbeta^\dag \rightarrow_p \bbeta_0$, $\hhat_{\bbeta^\dag} \rightarrow_p h_0$ and $E\left[\dot{\ell}(\bbeta_0, \bbeta^\dag, \hhat_{\bbeta^\dag})\right] = o_p(\|\bbeta^\dag - \bbeta_0\| + n^{-1/2}).$ Finally, suppose that there exists a neighborhood $\Wsc$ of $(\bbeta_0, \bbeta_0, h)$ such that $\{\dot{\ell}(t, \bbeta, h): (t, \bbeta, h) \in \Wsc\}$ is Donsker with square integrable envelope function and   $\{\ddot{\ell}(t, \bbeta, h): (t, \bbeta, h) \in \Wsc\}$ is Glivenko-Cantelli and bounded in $L_1$.

\section{Proof of sparsistency and asymptotic normality} \label{asymp}
To state and prove the results, we will need some preliminaries. Our objective function can be written equivalently as solely a function of $\bbeta$ rather than as a function of $\alphamj$ and $\dj$, as shown in Theorem 1 of \cite{zhou2010group}. For a fixed number of outcomes and fixed number of predictors, the objective function can be written
\begin{align}
	Q(\bbeta) = \| \bYtilde - \Xbbtilde\bbeta\|_2^2 + n\lambda_n\sumjp\Big\{\summM w_{j}\sm|\betamj|\Big\}^{1/2}\label{penlik2}
\end{align}
\noindent where $n\lambda_n = \sqrt{\lambda}$. In the following we establish the sparsistency and asymptotic normality of the minimizer $\bbetahat$. 

\subsection{Root-$n$ consistency} \label{consistency} 
We first show the root-$n$ consistency of our estimator $\bbetahat$. For PLLs \[
\{\Lscm(\bbetam)\}_{ m = 1, ..., M}\] that satisfy the regularity conditions listed above, if $\lambda_n = O_p(n^{-1/2})$, then there exists a local maximizer $\bbetahat$ of $Q(\bbeta)$ such that $\|\bbetahat - \bbeta_0\| = O_p(n^{-1/2}) .$

To see this, let $Q(\bbeta) =  \summM (\bbeta\sm - \bbetatilde\sm)\trans\bItilde\sm(\bbeta\sm-\bbetatilde\sm) + p_{\lambda_n,\bw}(\bbeta)$. We will show that for a given $\tau > 0$, $c = \min_{j,m}\{|\beta_{0j}\sm| : \beta_{0j}\sm \neq 0\}$ there exists a constant C such that
$P[\sup_{\|\bu \| = C} Q(\bbeta_0 + n^{-1/2} \bu) > Q(\bbeta_0)] \geq 1 - \tau$.
Now consider
\begin{align*}
	D(\bu) &=  Q(\bbeta_0 + n^{-1/2}\bu) - Q(\bbeta_0)
	= \summM (\bbeta_0\sm + n^{-1/2}\bu -\bbetatilde\sm)\trans\bItilde\sm(\bbeta_0\sm + n^{-1/2}\bu -\bbetatilde\sm)\\
	&  \qquad -  \summM (\bbeta_0\sm-\bbetatilde\sm)\trans\bItilde\sm(\bbeta_0\sm-\bbetatilde\sm) +
	 n\left(p_{\lambda_n, \bw}(|\bbeta_0 + n^{-1/2}\bu|) - p_{\lambda_n, \bw}(|\bbeta_0|)\right)\\
	&= n^{-1/2}\bu\trans\bItilde(\bbeta_0 - \bbetatilde) + \frac{n^{-1}}{2}\bu\trans\bItilde\bu - n\left(p_{\lambda_n, \bw}(|\bbeta_0 + n^{-1/2}\bu|) + p_{\lambda_n, \bw}(|\bbeta_0|)\right)\\
	&= (I) + (II) + (III)
\end{align*}
Now, since $\|\bItilde\sm - n\bI\sm\| = o_p(1)$, then $\|\bItilde - n\bI\| = o_p(1)$, and
$(I) = \nhalf\bu\trans \bI(\bbeta_0 - \bbetatilde)[1 + o_p(1)] 	\leq O_p(1)\|\bu\|\|\bI\|$.
Furthermore, $(II) = \bu\trans \bI\bu[1 + o_p(1)]
	\leq O_p(1)\|\bu\|^2\|\bI\|$.
Now, following the argument in \cite{zhou2010group},
$(III) \leq O_p(\lamn \nhalf)$.
Thus, as long as $\lamn = O_p(n^{-1/2})$, all terms are dominated by the first term of $(II)$, which is positive. And root-$n$ consistency follows.

\subsection{Sparsistency}
 We will now show that $\bbetahat$ is sparsistent: $P\left(\betahat_{ \Asc^c} = \bzero\right) \rightarrow 1.$ If we can show that
$\frac{\partial Q(\bbeta)}{\partial \beta_j\sm} = O_p(\nhalf) + n\frac{\partial p_{\lamn, \bw}(\bbeta)}{\partial \beta_j\sm}$
 then sparsistency follows from root-$n$ consistency and the argument in the proof of Theorem 4 in \cite{zhou2010group}. To this end, note that
\begin{align*}
	\frac{\partial Q(\bbeta)}{\partial \beta_j\sm} &= (\bbeta\sm - \bbetatilde\sm)\trans\bItilde_j\sm + n\frac{\partial p_{\lamn, \bw}(\bbeta)}{\partial \beta_j\sm}\\
	&= (\bbeta_0\sm - \bbetatilde\sm)\trans\bItilde_j\sm + (\bbeta\sm - \bbeta_0\sm)\bItilde_j\sm + n\frac{\partial p_{\lamn, \bw}(\bbeta)}{\partial \beta_j\sm}
	= O_p(\nhalf) + n\frac{\partial p_{\lamn, \bw}(\bbeta)}{\partial \beta_j\sm}
\end{align*}
for any $\bbeta$ satisfying $\|\bbeta - \bbeta_0\| = O_p(n^{-1/2})$, noting that $\bItilde_j\sm = O_p(n)$ for all $j, m$. And thus sparsistiency $P(\betahat_{ \Asc^c} = \bzero) \rightarrow 1$ follows.

\subsection{Asymptotic normality}
Next we consider asymptotic normality. Let $\bbeta(\Asc)$ denote $\bbeta$ with elements not in $\Asc$ set to 0. Because we have sparsistency, $\bbetahat(\Asc)$ is a root-$n$ consistent minimizer of $Q(\bbeta)$, and $\nabla Q\{\bbetahat(\Asc)\} = o_p(1).$ Thus, minimizing $Q\{\bbeta(\Asc)\}$ is asymptotically equivalent to minimizing
$
 Q_\Asc(\bbeta_{\Asc}) = {\blue }(\bbeta_{\Asc}-\bbetatilde_{\Asc})\trans\bItilde_{\Asc,\Asc}(\bbeta_{\Asc}-\bbetatilde_{\Asc})
 - (\bbeta_{\Asc}-\bbetatilde_{\Asc})\trans\bItilde_{\Asc,\Asc^c}\bbetatilde_{\Asc^c} +
n p_{\lamn, \bw}(\bbeta_{\Asc})
$]
where $\bItilde_{\Omega_1,\Omega_2}$ denotes the submatrix of $\bItilde$ corresponding to rows in $\Omega_1$  and columns in $\Omega_2$.
It follows that
\begin{align*}
	o_p(1) = \nabla Q_{\Asc}(\bbetahat_{ \Asc})
	&= \bItilde_{\Asc,\Asc}(\bbetahat_{\Asc} - \bbetatilde_{\Asc}) -\bItilde_{\Asc,\Asc^c} \bbetatilde_{\Asc^c}+ \nabla np_{\lamn, \bw}(\bbetahat_{\Asc}) \\
	&= \bItilde_{\Asc,\Asc}(\bbetahat_{\Asc} - \bbeta_{0\Asc}) + \bItilde_{\Asc,\Asc}(\bbeta_{0\Asc} - \bbetatilde_{\Asc}) - \bItilde_{\Asc,\Asc^c} \bbetatilde_{\Asc^c}
	+ \nabla np_{\lamn, \bw}(\bbetahat_{\Asc})
\end{align*}
and hence
\begin{align*}
	\nhalf(\bbetahat_{\Asc} - \bbeta_{0\Asc})
	& = \nhalf(\bbetatilde_\Asc - \bbeta_{0\Asc})
	-\nhalf\bItilde_{\Asc,\Asc}^{-1}\bItilde_{\Asc,\Asc^c} \bbetatilde_{\Asc^c} + \nhalf(n\bItilde_{\Asc,\Asc}^{-1})\nabla p_{\lamn, \bw}(\bbetahat_{\Asc}) \\
	& = (n\bItilde_{\Asc,\Asc}^{-1}) \nnhalf\sumin \bvp_{i\Asc}(\betaz) + \nhalf(n\bItilde_{\Asc,\Asc}^{-1})\nabla p_{\lamn, \bw}(\bbetahat_{\Asc}) + o_p(1)
\end{align*}
This, together with the same argument as in the proof of Theorem 4 in \cite{zhou2010group}, $\nabla p_{\lamn, \bw}(\bbetahat_{ \Asc}) = o_p(\nnhalf)$, implies that
$n^{1/2}(\bbetahat_\Asc - \bbeta_{0\Asc}) = n^{-1/2}\bI_{\Asc,\Asc}\inv\sumin \bvp_{i\Asc}(\bbeta_0) + o_p(1).$

\section{Pertubration-resampling}\label{app_ptb}
In this section, we give details on the perturbation resampling procedure and establish its asymptotic properties.

\subsection{Procedure for generating perturbed $\bbetahat^*$} Let $\Gsc = (G_1, ..., G_n)\trans$ be a vector of iid positive random variables with $E(G_i) = 1$ and Var($G_i$) = 1, generated independently of the data. We obtain $\bbetatilde^*$ as the maximizer of $\sum_{m=1}^M \Lsc^{(m)*}(\bbeta^{(m)})$ or explicitly as
$\bbetatilde^* = \bbetatilde + \sumin \bItilde^{-1} \bvptilde_i(\bbetatilde) (G_i - 1)$
where $\Lsc^{(m)*}(\bbeta)$ is the profile likelihood corresponding to the perturbed non-parametric likelihood with the contribution of the $i$th subject weighted by $G_i$, $\bItilde$ is the observed information matrix for $\bbeta$ evaluated at $\bbetatilde$, and $\bvptilde_i(\bbeta)$ is the empirical estimate of the score function $\bvp_i(\bbeta)$. In the second step, we find $\bbetahat^*$ as the minimizer of
$Q^*(\bbeta) = \summM (\bbeta\sm-\bbetatilde\smstar)\trans\bItilde\sm(\bbeta\sm-\bbetatilde\smstar) + \sumjp d_j + \lambda \summM\sumjp w_j\smstar|\alpham_j|$
subject to $d_j \ge 0$, $w_j\smstar = |\betatilde_j\smstar|\inv$.

\subsection{Properties of resampled $\bbetahat^*$}\label{ptb_proof}
In this section, we will show that
for PLLs $\{\Lscm(\bbetam)\}_{ m = 1, ..., M}$ that satisfy the regularity conditions listed in appendix \ref{appint}, if $n\inv\sqrt{\lambda} = o_p(n^{-1/2})$, then there exists a local maximizer $\bbetahat^*$ of $Q^*(\bbeta)$ such that \begin{enumerate}
\item[(i)] 
(i) $\|\bbetahat^* - \bbeta_0\| = O_p(n^{-1/2}),$ 
\item[(ii)] 
(ii)$P\left(\bbetahat^*_{\Asc^c} = 0 | \Vbb\right) \rightarrow 1$ as $n \rightarrow \infty,$ 
\item[(iii)] 
(iii) $\nhalf\left(\bbetahat^*_\Asc - \bbetahat_\Asc\right) \mid \Vbb$ converges in distribution to $N(0, \bI_{\Asc\Asc}\inv\bSigma_{\Asc\Asc}\bI_{\Asc\Asc}\inv)$.
\end{enumerate}
Let $\Pbb^*$ be the measure generated by both $\Vbb$ and $\Gsc$. First, note that
\begin{align*}
	\left\|\bbetatilde^* - \betaz\right\| &= \left\|\bbetatilde + \sumin \bItilde^{-1} \bvptilde_i(\bbetatilde) (G_i - 1) - \betaz\right\|
	\leq \left\|\bbetatilde - \betaz\right\| + \left\|\sumin \bItilde^{-1} \bvptilde_i(\bbetatilde) (G_i - 1)\right\|\\
	&= O_{\Pbb^*}(n^{-1/2}) + \left\| \frac{1}{n}\sumin\left\{n\bItilde^{-1} \bvptilde_i(\bbetatilde)\right\}\left( G_i - 1\right)\right\|
\end{align*}
\noindent  Noting that $\Gsc$ is independent of $\Vbb$, $E[G_i - 1] = 0$, and $ E[\bI^{-1} \bvptilde_i(\bbetatilde)] < \infty$,
$\frac{1}{n}\sumin\{n\bItilde^{-1} \bvptilde_i(\bbetatilde)\}( G_i - 1)\rightarrow_{\Pbb^*} 0$ and
the perturbed initial estimate is also root-$n$ consistent: $\left\|\bbetatilde^* - \betaz\right\| = O_{\Pbb^*}(n^{-1/2}).$

Moreover, for the root-$n$ consistency proof of $\bbetahat$, the role of $\bbetatilde$ in $Q(\bbeta)$ is only that of a root-$n$ consistent initial estimate.
Inspection of the proof of $\bbetahat$'s root-$n$ consistency will show that the only fact about $\bbetatilde$ that we need is $\left\|\bbetatilde - \bbeta_0\right\| = O_p(n^{-1/2})$. Therefore, in just the same way, root-$n$ consistency of $\bbetatilde^*$ gives us root-$n$ consistency of $\bbetahat^*$: $\left\|\bbetahat^* - \betaz\right\| = O_{\Pbb^*}(n^{-1/2})$.
Now, sparsistency of $\bbetahat^*\mid \Vbb$ follows from a similar argument as for sparsistency of $\bbetahat$. Consider
\begin{align*}
	\left.\frac{\partial Q^*(\bbeta)}{\partial \beta_j\sm}\right| \Vbb &= \left.(\bbeta\sm - \bbetatilde\smstar)\trans\bItilde_j\sm + n\frac{\partial p_{\lamn, \bw^*}(\bbeta)}{\partial \beta_j\sm}\right| \Vbb
	= \left.O_{\Pbb^*}(\nhalf) + n\frac{\partial p_{\lamn, \bw}(\bbeta)}{\partial \beta_j\sm}\right|\Vbb
\end{align*}

\noindent for any $\bbeta$ satisfying $\|\bbeta - \bbeta_0\| = O_{\Pbb^*}(n^{-1/2})$, noting that $\sumin(G_i - 1) = O_{\Pbb^*}(\nhalf)$. And thus sparsistency follows: $P\left(\left.\betahat^*_{ \Asc^c} = \bzero\right|\Vbb\right) \rightarrow 1.$

Finally, following the logic in the proof of asymptotic normality of $\bbetahat$,
\begin{align*}
	\nhalf(\bbetahat^*_{\Asc} - \bbeta_{0\Asc}) &= \nhalf(\bbetatilde^*_\Asc - \bbeta_{0\Asc})
	-\nhalf\bItilde_{\Asc,\Asc}^{-1}\bItilde_{\Asc,\Asc^c} \bbetatilde^*_{\Asc^c} + o_{\Pbb^*}(1)
	= \nhalf\bK (\bbetatilde^* - \bbeta_0) + o_{\Pbb^*}(1)
\end{align*}
for $\bK = d_\Asc - \bItilde_{\Asc\Asc}\inv\bItilde_{\Asc\Asc^c}d_{\Asc^c}, d_\Asc\bbeta = \bbeta_{\Asc},$ and $d_{\Asc^c}\bbeta = \bbeta_{Asc^c}$. In the proof of asymptotic normality of $\bbetahat$, we showed that
$n^{1/2}\bK (\bbetatilde - \bbeta_0) = n^{-1/2}\bI_{\Asc\Asc}\inv\sumin\bvp_{i\Asc}(\bbeta_0) + o_{\Pbb^*}(1).$
Note that $\nhalf\bK (\bbetatilde - \bbeta_0) = \nhalf\bK\bItilde\inv\sumin\bvptilde_i(\bbetatilde) + o_{\Pbb^*}(1)$, which suggests
\begin{align*}
\nhalf(\bbetahat^*_{\Asc} - \bbeta_{0\Asc}) 
&= n^{-1/2}\bI_{\Asc\Asc}\inv\sumin\bvp_{i\Asc}(\bbeta_0)(G_i - 1)+ \nhalf\bK(\bbetatilde - \bbeta_0) + o_{\Pbb^*}(1)
\end{align*}
And recall from above that $n^{1/2}(\bbetahat_\Asc - \bbeta_{0\Asc}) = n^{-1/2}\bI_{\Asc,\Asc}\inv\sumin \bvp_{i\Asc}(\bbeta_0) + o_{\Pbb^*}(1) = \nhalf\bK(\bbetatilde - \bbeta_0) + o_{\Pbb^*}(1).$
Then, let $\bZ_i = n^{-1/2}\bI_{\Asc\Asc}\inv\bvp_{i\Asc}(\bbeta_0)(G_i - 1)$, so that $E[\bZ_i | \Vbb] = 0$ and $\cov[\bZ_i | \Vbb] = n^{-1}\bI_{\Asc\Asc}\inv\bvp_{i\Asc}\bvp_{i\Asc}\trans\bI_{\Asc\Asc}\inv \equiv \Gamma_i$.
Because $\sumin E[ \| \Gamma_i^{-1/2} \|_2^3| \Vbb] = o_{\Pbb^*}(1)$, then by the argument in \cite{bentkus2005lyapunov}, \[\left.\nhalf(\bbetahat^*_\Asc - \bbetahat_\Asc) \right| \Vbb \rightarrow_{\Lsc} N(0, \bI_{\Asc\Asc}\inv\bSigma_{\Asc\Asc}\bI_{\Asc\Asc}\inv).\]
\section{Testing}
\subsection{Justification of stepdown procedure}\label{step}
In this section, we will show that, when using an estimator that satisfies $P(\bbetahat_{\Asc^c} = 0) \rightarrow 1$, SMRT has asymptotic FWER of 0 for any reference distribution, $\psi$, and $k$. We will discuss controlling the FWER for a single predictor. Controlling FWER for all predictors follows the same logic but is notationally burdensome. We will first consider $k = 1$ and then discuss $k > 1$.

First, take $k = 1$. \cite{goeman2010sequential} show that two conditions need to be satisfied in order for a sequentially rejective procedure of this sort to control the FWER at a given level $\alpha$. First, a {\em monoticity} condition requires that the threshold for rejection must not increase as the test proceeds, and, even more, that for any $\Omega_k \supset \Omega_{k'}$ (even those not observable as part of the same stepdown test):
\begin{align}
c_j^{\Omega_k}(\psi) \geq c_j^{\Omega_{k'}}(\psi). \label{cond1}
\end{align}
This condition is guaranteed by construction. Recall that $c_j^{\Omega}(\psi)$ is the $\psi$th quantile of \[\{\max_{m \in {\Omega}} t_j^{*_b(m)}\}_{b = 1, .., B}\] and note that for each $b$, 
\begin{align*}
\max_{m \in {\Omega_k}} t_j^{*_b(m)} \geq \max_{m \in {\Omega_{k'}}} t_j^{*_b(m)}
\end{align*}
because $\Omega_{k'} \subset \Omega_k$. This in turn implies \eqref{cond1}.
Second, a {\em single-step} condition requires that the thresholds must be chosen so as to control type I error at $\alpha$ in the {\em critical case}, when the set of candidate hypotheses are all null. That is, let $\Rscr_{0j} \in \Hsc_j$ be the set of indices of all true null hypotheses, then for any $\Rscr_{0j}$,
$P\left(s_j^{\Rscr_{0j}} > c_j^{\Rscr_{0j}}(\psi)\right) \leq \alpha.$
Because $P(\bbetahat_{ \Asc^c} = 0) \rightarrow 1$, any choice of $\psi$ and reference distribution will be sufficient. That is, for any $\psi$ and reference distribution,
$P\left(s_j^{\Rscr_{0j}} > c_j^{\Rscr_{0j}}(\psi)\right) \rightarrow 0$ because $P(s_j^{\Rscr_{0j}} = 0) \rightarrow 1$ and $c_j^{\Rscr_{0j}}(\psi) \geq 0$ for any $\psi$ and any reference distribution.
Thus, our testing procedure will control the FWER for each predictor $x_j$ asymptotically at any level $\alpha$ for any choice of $\psi$ and any reference distribution. Since we can choose $\alpha$ as small as we want, the FWER for the set of hypotheses $\Hsc_j$ converges to 0.

When $k > 1$, recall that SMRT proceeds by first holding out the $k-1$ largest test statistics and performing a stepdown test on the remaining $M-k+1$. Then, if any hypotheses are rejected among the remaining $M-k+1$, all $k-1$ of the held out hypotheses are rejected. Thus, there are two sources of error: a common error, which can be incurred during the stepdown test, and a type I error by implication, which can occur by falsely rejecting one of the $k-1$ held out hypotheses. By following the argument for $k = 1$, we can see that the probability of making any common error decreases to 0 as $n \rightarrow \infty$. We now show that the probability of an error by implication also vanishes asymptotically. The probability of making an error by implication is \[
P\left[\bigcup_{m \in \mathcal{N}}\{t_j\sm \geq t_j^{(r_k)} > c_j^{\Omega_1}(\psi)\}\right]
\]
where $\mathcal{N} = \{r_l : l < k, H_j^{(r_l)}\}$ is the index set of the true nulls in the held out $k-1$ hypotheses. However, for every $m \in \mathcal{N}, t_j\sm \rightarrow 0$ as $n \rightarrow \infty$, regardless of $\psi$ and reference distribution, by the sparsistency of $\bbetahat$, which means that $P\left[\bigcup_{m \in \mathcal{N}}\{t_j\sm \geq t_j^{(r_k)} > c_j^{\Omega_1}(\psi)\}\right] \rightarrow 0$, and thus the FWER of SMRT when $k > 1$ asymptotically vanishes.

\subsection{Extending FWER control to all predictors}\label{allpredix}
To extend FWER control to the set of all hypotheses $\{H_j\sm\}_{j=1, ..., p; m = 1, .., M}$, one could test each $\Hsc_j$ at level $\alpha/p$. This may be a conservative strategy, as it relies on a union-bound argument. 
One could also easily extend the stepdown testing to the set of all test statistics $\bt = (\bt_1\trans, ..., \bt_p\trans)\trans$. If $k = 1$, then one can simply perform the procedure from section \ref{testing}
on $\bt$. If $k > 1$, then one would first identify and throw out the $k-1$ largest test statistics {\em for each predictor} to produce a new vector of test statistics $\bt_k$ which includes only the $k$ smallest test statistics for each predictor. Then we perform a stepdown test on $\bt_k$ to obtain a set of rejected hypotheses $\Rscr$. For each predictor that has any rejected hypotheses in $\Rscr$, we also reject the hypotheses corresponding to that predictor's largest $k-1$ test statistics as well. So, again, for each predictor, we would reject either 0 or greater than or equal to $k$ hypotheses while controlling the FWER for all predictors simultaneously. 

\section{Algorithm}\label{algorithm}
An iterative procedure can be employed to fit the model \eqref{penlik1}. First, fix $\bd$ and estimate $\balpha$ via adaptive lasso. Next, fix $\balpha$ and estimate $\bd$ using the nonnegative garrote. However, because of the widespread availability and speed of lasso-type estimation, we in general prefer to employ adaptive lasso to the nonnegative garrote. So we propose to estimate $\bd$ using adaptive lasso as well, by minimizing the following objective function

\begin{align}
& \| \bYtilde - \Xbbtilde\bbeta\|_2^2 + \sumjp |d_j| + \lambda \summM\sumjp w_j\sm| \alpham_j|, \label{penlik3}
\end{align}

\noindent Using the adaptive lasso in place of the nonnegative garrote is justified here by the argument in \cite{zou2006adaptive} that the adaptive lasso is asymptotically equivalent to the nonnegative garrote. That is, the nonnegative garrote is equivalent to the adaptive lasso with a further sign constraint, and the sign constraint is satisfied (at least in the limit) by consistency of the adaptive lasso. Our iterative fitting procedure, then, uses the adaptive lasso at both stages and is thus very fast.

Fitting the model can proceed as follows:\\
1. Set $\bd_{(0)} = \bone$ and $\Xbbtilde_\beta = \Xbbtilde \text{diag}(|\bbetatilde|)$. Let $k$ = 1.\\
2. Update $\balpha$. Set $D_\alpha = \text{diag}(\bd_{(k-1)})$ and $\Xbbtilde_\alpha = \Xbbtilde_\beta \text{diag}(D_\alpha, ..., D_\alpha)_{Mp \times Mp}$
and obtain
\begin{align*}
	\balpha_{(k)} = \argmin{\balpha}{\| \bYtilde - \Xbbtilde_\alpha\balpha\|_2^2 + \lambda \summM\sumjp| \alpham_j|}
\end{align*}
Inclusion of $\bbetatilde$ in $\Xbbtilde_\alpha$ is equivalent to using weights $w_j\sm = |\betatilde_j\sm|^{-1}$.

\noindent 3. Update $\bd$. Set $\Xbbtilde_d = \Xbbtilde \bA_d$ where $\bA_d = \left[
\begin{matrix}
\text{diag}(\balpha\sone_{(k)})_{p\times p}\\
\text{diag}(\balpha\stwo_{(k)})_{p\times p}\\
\vdots\\
\text{diag}(\balpha\sM_{(k)})_{p\times p}
\end{matrix}\right]_{Mp \times p}$ \\
Then,
\begin{align*}
	\bd_{(k)} = \argmin{\bd}{\| \bYtilde - \Xbbtilde_d\bd\|_2^2 + \sumjp | d_j|}
\end{align*}

\noindent 4. Update $\bbeta$. $$\beta_{j(k)}\sm = d_{j(k)}\alpha_{j(k)}\sm|\betatilde_j\sm|$$.\\
5. Iterate until convergence.

\section{Reference distribution details}\label{app_ref}
In this section, we discuss in more detail the choice and generation of reference distributions. Ideally, one would choose a reference distribution that approximates the finite-sample distribution of the test statistic $t_j\sm$ under $H_j\sm$ and maintains the correlation structure across the $M$ outcomes. We explore some possibilities below.

An immediately appealing choice for the reference distribution is to use the resampled $\bbetahat^*$, since, as we stated in section \ref{sec-estimation}, $\nhalf\left(\bbetahat^* - \bbetahat\right) \mid \Vbb$ has the same asymptotic distribution as $\nhalf\left(\bbetahat - \bbeta_0\right)$. Furthermore, resampling maintains the correlation structure across the outcomes. 
Thus, we could choose $t_j^{*_b(m)} = \nhalf\left|\betahat_j^{*_b(m)} - \betahat_j\sm\right|/\sigmatilde_j\sm$ with the expectation that (for $n$ large enough) $\Tsc_j\sm$ will approximate well the finite-sample distribution of $t_j\sm$. 
Simulation results suggest that, although resampling provides good approximation to the finite-sample distribution of $\bbetahat_{\Asc}$, it tends to over-estimate the variability of $\bbetahat_{\Asc^c}$ (see figure \ref{ptbpmt}). By taking advantage of the sparsity in the estimator, however, one can maintain a desired FWER ($\alpha$) while achieving high power by setting $\psi$ at a less conservative level ($\psi < 1 - \alpha$). We demonstrate the promise of this reference distribution in figure \ref{ptb_rej_plot}. However, empirically identifying a proper $\psi$ to both preserve the FWER and achieve high power may be computationally challenging in practice.

Alternatively, one may use permutations to obtain a better finite-sample approximation of the distribution of $t_j\sm$.
Let $\bbetahat(\Omega) = \{\betahat_j\sm(\Omega)\}_{j = 1, ..., p; m = 1, ..., M}$ denote the estimate of $\bbeta_0$ using the dataset $\{(\by_i^{\Omega\trans},\bx_i\trans)\trans\}_{ i = 1, ..., n}$, where $\{\by_i^{\Omega}\}_{i=1,...,n}$ denotes a partially permuted counterpart of $\{\by_i\}_{i=1,...n}$ with $\{y_i\sm\}_{m \in \Omega;i=1,...,n}$ randomly permuted across subjects but $\{y_i\sm\}_{m \notin \Omega;i=1,...,n}$ unchanged. And let $\bbetahat^{_b}(\Omega)$ be the $b$th such permutation-based estimate. To be clear, for example, $\bbetahat^b(\{1\})$ corresponds to the estimate of $\bbeta_0$ from a dataset where only the first outcome $\{y_i\sone\}_{i=1, ..., n}$ is permuted, and $\bbetahat^b(\{1, 3\})$ corresponds to the estimate of $\bbeta_0$ from a dataset where both the first outcome and third outcomes $\{(y_i\sone, y_i^{(3)})\}_{i=1, ..., n}$ are permuted.

A reference distribution that we pursue in our simulations is a composite distribution obtained by permuting each of the outcomes individually. For each $m$, we obtain $\left\{\bbetahat^{_b}(\{m\})\right\}_{b=1, .., B}$ and retain only those elements which pertain to outcome $m$: $\left\{\betahat_j^{_b(m)}(\{m\})\right\}_{j=1, ..., p; b = 1,..., B}$. We then define the reference distribution for the stepdown procedure as $t_j^{*_b(m)} = \nhalf\left|\betahat_j^{_b(m)}(\{m\})\right|/\sigmatilde_j\sm.$
In this way, we are essentially obtaining a reference distribution for $t_j\sm$ under the null hypothesis $\bigcap_{j = 1, ..., p}H_j\sm$. This strategy has the undesirable consequence of breaking the correlation structure across outcomes, since $t_j^{*_b(m)}$ and $t_j^{*_b(m')}$ are obtained under different permutation regimes for $m \neq m'$. But defining $t_j^{*_b(m)}$ in this way allows us to approximate the distribution of $t_j\sm$ without making any assumption about $\{H_j^{(m')}\}_{m' \neq m}$. Not making any assumption about $\{H_j^{(m')}\}_{m' \neq m}$ is important because, for example, the null distribution of $t_j\sm$ will be very different under $\bigcap_{m' \neq m} H_j^{(m')}$ as opposed to under $\bigcap_{m' \neq m} \Hbar_j^{(m')}$ because of the group penalty in the penalized likelhood \eqref{garrote}.

\bibliography{bib_two}
\bibliographystyle{plain}

\label{lastpage}
\end{document}

%% file: GrandMacros.tex
\def\bzero{{\bf 0}}
\def\bone{{\bf 1}}
\def\bb{{\bf b}}

\def\bd{{\bf d}}

\def\bt{{\bf t}}
\def\bu{{\bf u}}

\def\bw{{\bf w}}
\def\bx{{\bf x}}
\def\by{{\bf y}}
\def\bz{{\bf z}}
\def\bA{{\bf A}}

\def\bI{{\bf I}}

\def\bK{{\bf K}}

\def\bV{{\bf V}}

\def\bY{{\bf Y}}
\def\bZ{{\bf Z}}

\def\thick#1{\hbox{\rlap{$#1$}\kern0.25pt\rlap{$#1$}\kern0.25pt$#1$}}
\def\balpha{\boldsymbol{\alpha}}
\def\bbeta{\boldsymbol{\beta}}

\def\bepsilon{\boldsymbol{\epsilon}}

\def\bLambda{{\boldsymbol{\Lambda}}}

\def\bSigma{\boldsymbol{\Sigma}}

\def\smbalpha{\boldsymbol{{\scriptstyle{\alpha}}}}

\def\hhat{{\widehat h}}

\def\bItilde{{\widetilde \bI}}

\def\bYtilde{{\widetilde \bY}}

\def\betahat{{\widehat\beta}}

\def\sigmahat{{\widehat\sigma}}

\def\betatilde{{\widetilde\beta}}

\def\sigmatilde{{\widetilde\sigma}}

\def\bbetahat{{\widehat\bbeta}}

\def\smbalpha{\widehat{\smbalpha}}

\def\bbetatilde{{\widetilde\bbeta}}

\def\bLambdatilde{{\widetilde\bLambda}}

\def\hbar{\bar{ h}}

\def\Hbar{\bar{ H}}

\def\Asc{{\cal A}}
\def\Bsc{{\cal B}}

\def\Gsc{{\cal G}}
\def\Hsc{{\cal H}}

\def\Lsc{{\cal L}}

\def\Rsc{{\cal R}}
\def\Ssc{{\cal S}}
\def\Tsc{{\cal T}}

\def\Wsc{{\cal W}}

\def\Zsc{{\cal Z}}

\def\transpose{{\sf \scriptscriptstyle{T}}}

\def\half{\frac{1}{2}}

\def\nhalf{n^{\half}}

\def\nnhalf{n^{-\half}}

\def\var{\mbox{var}}

\def\sumin{\sum_{i=1}^n}

\def\trans{^{\transpose}}
\def\inv{^{-1}}

\def\argmindum{\mathop{\mbox{argmin}}}
\def\argmin#1{\argmindum_{#1}}
\def\argmaxdum{\mathop{\mbox{argmax}}}
\def\argmax#1{\argmaxdum_{#1}}

\def\cov{\mbox{cov}}
\def\diag{\mbox{diag}}

\def\diag{\mbox{diag}}

\def\var{\mbox{var}}

\def\mybox#1{\vskip1mm \begin{center}
        \hspace{.0\textwidth}\vbox{\hrule\hbox{\vrule\kern6pt
\parbox{.9\textwidth}{\kern6pt#1\vskip6pt}\kern6pt\vrule}\hrule}
        \end{center} \vskip-5mm}
\def\lboxit#1{\vbox{\hrule\hbox{\vrule\kern6pt
      \vbox{\kern6pt#1\vskip6pt}\kern6pt\vrule}\hrule}}

\def\thickboxit#1{\vbox{{\hrule height 1mm}\hbox{{\vrule width 1mm}\kern6pt
          \vbox{\kern6pt#1\kern6pt}\kern6pt{\vrule width 1mm}}
               {\hrule height 1mm}}}

\def\fat#1{\hbox{\rlap{$#1$}\kern0.25pt\rlap{$#1$}\kern0.25pt$#1$}}